\documentclass[12pt,a4paper]{article}

\usepackage[english]{babel}

\usepackage{indentfirst}
\usepackage{amsmath}
\usepackage{amssymb}
\usepackage{soul}
\usepackage[pdftex]{graphicx,color}

\definecolor{green}{rgb}{0.0,0.6,0.0}

\usepackage{geometry}
\begin{document}

\begin{flushright}
UVa/270719
\\
%\today
\end{flushright}
\bigskip

%\hfill \today

\begin{center}
\textbf{\LARGE \boldmath 
Acceleration of an unpolarized proton along a uniform magnetic field: Casimir momentum of leptons}

\vspace*{1.6cm}

{\large Manuel Donaire}\footnote{manuel.donaire@uva.es}

\vspace*{0.4cm}

\textsl{%
Departamento de F\'isica Te\'orica, At\'omica y \'Optica and IMUVA,  Universidad de Valladolid, \\Paseo Bel\'en 7, 47011 Valladolid, Spain\\[5mm]
}
\vspace*{0.8cm}

\textbf{\Large Abstract}\\[10pt]
\parbox[t]{0.9\textwidth}{It has been recently shown that a chiral molecule accelerates linearly along a spatially uniform magnetic field, as a result of the parity-time symmetry breaking induced in its QED self-interaction. In this work we extend this result to fundamental particles which present EW self-interaction, in which case parity is violated by the EW interaction itself. In particular, we demonstrate that, in a spatially uniform and adiabatically time-varying magnetic field, an unpolarized proton coupled to the leptonic vacuum acquires a kinetic momentum antiparallel to the magnetic field, whereas virtual leptons gain an equivalent \emph{Casimir momentum} in the opposite direction. That momentum is proportional to the magnetic field and to the square of Fermi's constant. We prove that the kinetic energy of the proton is a magnetic energy which forms part of its EW self-energy.
 
 }

\end{center}

\vspace*{3cm}
\newpage

\section{Introduction}\label{sec:intro}

The vacuum state of any quantum field theory contains field fluctuations whose energy is commonly referred to as zero-point energy \cite{Miltonbook,Milonnibook}. Such an energy, which is conjectured to form part of the cosmological constant \cite{Weinberg}, could have observable gravitational effects \cite{Jaffe}. In addition, when quantum field fluctuations couple to material current fluctuations, their interaction energy is finite and can be detected through physical observables of the material degrees of freedom. That is, for instance, through  the forces between conducting plates in the Casimir effect \cite{Casimir}, the van-der-Waals forces between neutral polarizable molecules \cite{vdW}, or the Lamb-shift of neutral atoms \cite{Bethe,Milonnibook}. Beyond QED,  analogous Casimir effects have been addressed in QCD and non-Abelian gauge theories \cite{cavitychromo,Cherenkov}, as well as in braneworld scenarios \cite{PRL99071601(2007)Durrer}.

On the other hand, homogeneity of space and translation invariance of the vacuum state of any quantum field theory imply that the net momentum of its vacuum field fluctuations is identically zero. However, when quantum field fluctuations couple to material current fluctuations through a Hamiltonian which violates both parity (P) and time-reversal (T), it is symmetry allowed for the field fluctuations to acquire a net virtual momentum.  Conversely, in virtue of total momentum conservation, the material degrees of freedom gain a kinetic momentum of equal strength in the opposite direction, which can be experimentally accessible.

The existence of a virtual momentum of electromagnetic (EM) fluctuations was first shown by Feigel in ref.~\cite{Feigel} --see also ref.~\cite{Croze}. There, P and T symmetries appeared broken by the magneto-optical activity of a chiral medium in a magnetic field. Feigel obtained a net momentum for the fluctuations of the effective EM field, which he referred to as  \emph{Casimir momentum} for its analogy with the energy of the EM fluctuations in the Casimir effect. However, Feigel's semiclassical approach, based on the application of the fluctuation-dissipation theorem to an effective medium, contained unphysical divergences. Later, the QED approach of ref.~\cite{PRLDonaire} revealed that a finite Casimir momentum is indeed carried by the virtual photons which couple to a chiral molecule in the presence of a magnetic field. In contrast to Feigel's, the approach of ref.~\cite{PRLDonaire} was Hamiltonian, with P being broken by the electrostatic potential generated by the chiral distribution of atoms within the molecule, and T being broken by the Zeeman potential of the magnetic field. Further, it was shown that the kinetic energy of the molecule is a Doppler-shift correction to the QED self-energy of a chromophoric electron within the molecule \cite{JPCMDonaire}.

The main purpose of this article is to demonstrate that the Casimir momentum is not an effect unique to photons interacting with chiral molecules, but a generic phenomenon in nature. This is so because the electroweak (EW) interaction of fundamental particles violates parity-symmetry by itself. This implies that any fundamental particle can be accelerated along spatially uniform magnetic fields. In order to demonstrate these facts, we compute the Casimir momentum of the virtual leptons which couple electroweakly to an actual unpolarized proton in the presence of a magnetic field. %In contrast to EM, parity is not a symmetry of the electroweak (EW) interaction, which is chiral. 
 The underlying mechanism through which the proton accelerates is analogous to that which explains the directionality of beta emission in the decay of polarized nuclei \cite{Wu}, but for the fact that while leptons are actual particles in the latter case, they remain virtual in the former one. That is, in the first place chirality in the EW interaction implies an alignment between the spins and the momenta of the hadrons and leptons which participate in the EW self-interaction of the proton. %In particular, the orientation of the spin of any particle which participates in the EW self-interaction of the actual proton induces a net momentum on it,
   On the other hand,  the magnetic particles get partially polarized along the magnetic field, which breaks time-reversal. Hence, virtual leptons are emitted and absorbed by the proton with different probabilities in the directions parallel and antiparallel to the magnetic field, resulting in a net transfer of momentum to the actual proton, as well as to the virtual leptons, along the polarization axis.  Ultimately, we aim at unraveling the transfer of momentum and energy among the proton, the EW vacuum and the magnetic field. In this respect, we show that whereas the kinetic energy gained by the proton is a Doppler-shift correction to its EW self-energy, which is provided by the source of magnetic field, its momentum is provided by virtual leptons. 
 
 The article is organised as follows. In section \ref{Approach} we describe the fundamentals of our approach and the Hamiltonians of the relevant interactions. In section \ref{Adiabatic} we compute the lepton Casimir momentum as well as the velocity reached by a proton along an external magnetic field. The nature of the proton kinetic energy is explained in section \ref{AdiaEnergy}.  The conclusions are summarized in section \ref{Conclusions} together with some relevant comments.

\section{Fundamentals of the approach}\label{Approach}

We adopt a time-dependent quantum perturbative approach, using the Hamiltonian formalism in Schr\"odinger's representation. Slowly time-varying quantities like momenta and energies are evaluated in the adiabatic limit at zero temperature. As for the Hamiltonian, we use the Dirac Hamiltonian for free nucleons and leptons in an adiabatic magnetic field, a low-energy effective  potential for their EW interaction, and a simplified confining Hamiltonian to model the internal dynamics of the constituent quarks within the nucleons. Therefore, while nucleons and leptons are described as low-energy quantum fields, the quarks within the nucleons are treated as quantum mechanical states.
 
Let us denote by $H=H_{0}+W$ the total Hamiltonian of the system in a spatially uniform magnetic field $\mathbf{B}$, where  $H_{0}=\sum_{\alpha}H^{\alpha}_{D}+H^{\alpha}_{mag}+H_{int}^{N}$ is the nonperturbative Hamiltonian, and $W$ is the perturbative EW interaction. In $H_{0}$, $H^{\alpha}_{D}$ is the Dirac Hamiltonian of free particles of the kind $\alpha$, while  $H^{\alpha}_{mag}$ accounts for their interaction with $\mathbf{B}$. The particles are fundamental leptons, electrons ($e$) and neutrinos ($\nu$); and composite nucleons, protons ($p$) and neutrons ($n$), made of up $(u)$ and down $(d)$ quarks.  Regarding nucleons, the dynamics of their internal and external degrees of freedom are considered decoupled. %\cite{remark}. 
As for their external dynamics, protons and neutrons are Dirac fields which may be considered the two components of an isospin doublet of isospin $I=1/2$, approximately equal masses, and third isospin components $I_{3}= \pm1/2$, respectively. Hereafter we will label with a script $N$ all those operators which act upon the nucleon doublet.  
 As for the internal dynamics, we consider up and down quarks as constituent quarks with equal masses $m_{d}\approx m_{u}$, and treat them as quantum mechanical states. They are bound within the nucleons by some effective strong confining potential which, together with their internal kinetic energy, conform the internal nucleon Hamiltonian $H_{int}^{N}$. Lastly,  $W$ is the low-energy four-fermion EW-Hamiltonian of interaction between nucleon and lepton fields.

\subsection{Dirac Hamiltonian with an adiabatic magnetic field: Landau levels}\label{hamil}

As for the Dirac and magnetic Hamiltonians in $H_{0}$, they read in natural units, $\hbar=c=1$,
\begin{align}
H^{\alpha}_{D}&=\int\textrm{d}^{3}R\:\bar{\Psi}_{\alpha}(-i\gamma^{j}\partial_{j}+m_{\alpha})\Psi_{\alpha},\label{HD}\\ 
H^{\alpha}_{mag}&=\int\textrm{d}^{3}R\bar{\Psi}_{\alpha}[-q_{\alpha}\gamma^{j}\tilde{A}_{j}-\frac{\kappa_{\alpha}\mu_{N}}{2}S^{ij}\tilde{F}_{ij}]\Psi_{\alpha},\label{Hmag}
\end{align}
where $i,j=1,2,3$ are spatial indices, and the script $\alpha$ denotes the kind of particle, $\alpha=p,n,e,\nu$. $\Psi_{\alpha}$ are Dirac spinors with charges $q_{\alpha}$ and masses $m_{\alpha}$, and  spatial derivatives for nucleons are with respect to their center of mass, $\mathbf{R}$. We consider that each nucleon interacts with the magnetic field as a whole \cite{Karliner}, with $\kappa_{p,n}\mu_{N}/2$ being their anomalous magnetic moments, $\kappa_{p,n}/2$ the anomaly factors, and $\mu_{N}$ the nuclear magneton \cite{Scadron,Mao}. $S_{ij}=i[\gamma_{i},\gamma_{j}]/4$ is proportional to the spin operators \cite{Peskin}, $\tilde{F}_{ij}=\partial_{i}\tilde{A}_{j}-\partial_{j}\tilde{A}_{i}$ is the electromagnetic tensor, and our gauge choice for the electromagnetic vector, $\tilde{\mathbf{A}}=-By\hat{\mathbf{x}}$, yields an external magnetic field $\mathbf{B}$ along the $\hat{\mathbf{z}}$-axis, which increases adiabatically from $\mathbf{0}$ to a final value $\mathbf{B}_{0}$. Thus, $H^{\alpha}_{mag}$ evolves adiabatically with $\mathbf{B}$ until $\mathbf{B}=\mathbf{B}_{0}$, and remains constant from them on. Under these conditions, the eigenstates of $H_{0}(\mathbf{B})$ evolve in time with $\mathbf{B}$ towards the \emph{corresponding} eigenstates of the stationary Hamiltonian $H_{0}(\mathbf{B}_{0})$, with $\mathbf{B}$ considered a quasi-stationary parameter in $H_{0}(\mathbf{B})$ at any time. As explained in appendix \ref{Dyna}, we will replace the time-dependence of the eigenstates of $H_{0}(\mathbf{B})$ with their dependence on $\mathbf{B}$, which will appear as a subscript. Lastly, the value of $B$ is such that the inequality $eB\ll m_{d}^{2}$ is assumed throughout this article. 

The eigenstates of $H^{e}_{D}+H^{e}_{mag}(\mathbf{B})$ and $H^{p}_{D}+H^{p}_{mag}(\mathbf{B})$, for electrons and protons respectively, correspond to Landau levels. In our gauge, they are characterized by two quantum numbers, principal number $\mathcal{N}=0,1,2,...$ and spin number $\mathcal{S}=\pm1$; and by the continuous components of their canonical conjugate momenta along $\hat{\mathbf{x}}$ and $\hat{\mathbf{z}}$, $\mathbf{p}$ and $\mathbf{k}$ for protons and electrons, respectively, $|\mathcal{N},\mathcal{S};\mathbf{p}(\mathbf{k})\rangle^{p,e}_{\mathbf{B}}$. 
 Virtual electrons and positrons are field fluctuations of the leptonic vacuum, $|\Omega_{l}\rangle_{\mathbf{B}}$, whose energies depend only on $\mathcal{N}$ according to $E^{e}_{\mathcal{N}}=\sqrt{m_{e}^{2}+k_{z}^{2}+2\mathcal{N}eB}$. In terms of creation operators for positrons, $b_{\mathcal{N},\mathcal{S}}^{e\dagger}$, the positron eigenstates read  %and, except for $N=0$, are $two-fold degenerate. 
$|\mathcal{N},\mathcal{S};k_{x},k_{z}\rangle_{\mathbf{B}}^{e}=\sqrt{2E_{\mathcal{N}}^{e}}\:b_{\mathcal{N},\mathcal{S}}^{e\dagger}(k_{x},k_{z})|\Omega_{l}\rangle_{\mathbf{B}}$.
 As for the actual proton, its state is at any time a linear combination of the two lowest Landau levels with $\mathcal{N}=0$ and $\mathcal{S}=\pm1$, $|0,1;\mathbf{p}\rangle^{p}_{\mathbf{B}}$ and $|0,-1;\mathbf{p}\rangle^{p}_{\mathbf{B}}$, with energies  $E^{p}_{\pm}=\sqrt{m_{p}^{2}+p_{z}^{2}+(1\mp1)eB}\mp \kappa_{p}\mu_{N}B/2$, respectively. %$E^{p}_{S}=[(\sqrt{m_{p}^{2}+(1-S)eB}-Sg_{p}\mu_{N}B/2)^{2}+p_{z}^{2}]^{1/2}$ 
In terms of creation operators for protons, $a^{p\dagger}_{\mathcal{N},\mathcal{S}}$, acting upon the hadronic vacuum $|\Omega_{h}\rangle_{\mathbf{B}}$, the eigenstates of the actual proton read $|0,\pm1;\mathbf{p}\rangle^{p}_{\mathbf{B}}=\sqrt{2E_{\pm}^{p}}a^{p\dagger}_{0,\pm1}(p_{x},p_{z})|\Omega_{h}\rangle_{\mathbf{B}}$.

Neutron states of momentum $\mathbf{p}$ experience a Zeeman splitting which depends only on the spin number $\mathcal{S}=\pm1$, $E^{n}_{\pm}=[(\sqrt{m_{n}^{2}+p_{x}^{2}+p_{y}^{2}}\mp \kappa_{n}\mu_{N}B/2)^{2}+p_{z}^{2}]^{1/2}$. Finally, massless neutrinos are free fermions of momentum $\mathbf{q}$ and energy  $E_{q}=q$.  

Expressions of the spinor eigenstates as well as of the fermion fields of all the species are compiled in appendix \ref{AppA} --cf. \cite{Bhattacharya,Bander,Mao,Broderick}.

\subsection{Internal nucleon effective Hamiltonian, $H_{int}^{N}$}\label{Isgur}

For the sake of simplicity we adopt the simplest effective quantum mechanical model to describe the strong confinement of quarks within the nucleons. Such a model consists of considering the three quarks, up and down quarks, as constituent quarks of equal masses, $m_{u}\approx m_{d}$, whose dynamics is nonrelativistic. They are bound together by a strong confining potential which is modelled by a harmonic-oscillator potential, $V_{\textrm{conf}}^{N}$. This is a simplification of the models of refs.~\cite{Coince,IsgurKarl1978,IsgurKarl1979,deRujula}, from which we disregard higher order corrections associated to Coulomb and hyperfine interactions between quarks. 

The relevant EW interaction in our calculations takes place within the nucleon at the location $\tilde{\mathbf{r}}$ of an active quark, up for proton, down for neutron, while the remaining quarks rest as spectators [see figure~\ref{fig1}]. We denote the position vectors of the three constituent quarks of the nucleon by $\mathbf{r}_{u}$,  $\mathbf{r}_{d}$, and $\tilde{\mathbf{r}}$ ,where $\mathbf{r}_{u}$ and $\mathbf{r}_{d}$ refer to the up and down spectator quarks, and $\tilde{\mathbf{r}}$ is the position vector of the active quark in the laboratory frame. Their corresponding conjugate momenta are $\mathbf{p}_{u}$,  $\mathbf{p}_{d}$, and $\tilde{\mathbf{p}}$.  The Jacobi position vectors and corresponding conjugate momenta read \cite{Coince}, 
\begin{align}
\mathbf{R}&=\frac{\mathbf{r}_{u}+\mathbf{r}_{d}+\tilde{\mathbf{r}}}{3},\quad\mathbf{r}_{\rho}=\mathbf{r}_{d}-\mathbf{r}_{u},\quad\mathbf{r}_{\lambda}=\tilde{\mathbf{r}}-\frac{\mathbf{r}_{u}+\mathbf{r}_{d}}{2},\nonumber\\
\mathbf{p}&=\mathbf{p}_{u}+\mathbf{p}_{d}+\tilde{\mathbf{p}},\:\:\mathbf{p}_{\rho}=\frac{\mathbf{p}_{d}-\mathbf{p}_{u}}{2},\:\:\mathbf{p}_{\lambda}=\frac{2\tilde{\mathbf{p}}-\mathbf{p}_{u}-\mathbf{p}_{d}}{3},\nonumber
\end{align}
where $\mathbf{R}$ is the centre of mass of the nucleon and $\mathbf{p}$ its conjugate momentum.  As for the active quark, its position vector  $\tilde{\mathbf{r}}$ reads, in terms of Jacobi's coordinates, $\tilde{\mathbf{r}}=\mathbf{R}+2\mathbf{r}_{\lambda}/3$. In the nonrelativistic limit, the total kinetic energy of the nucleon is, in terms of Jacobi's momenta, 
\begin{align}
\frac{p^{2}_{u}+p_{d}^{2}+\tilde{p}^{2}}{2m_{d}}+\mathcal{O}(eB/m_{d})=\frac{p^{2}/6+p_{\rho}^{2}+3p_{\lambda}^{2}/4}{m_{d}}+\mathcal{O}(eB/m_{d}),\label{kiny}
\end{align}
from which we identify the first term on the right hand side with the kinetic energy of the centre of mass, already accounted for in $H_{D}^{p,n}$; and  the reminder, $p_{\rho}^{2}/m_{d}+3p_{\lambda}^{2}/4m_{d}$, with the internal kinetic energy of the quarks within the internal nucleon Hamiltonian, $H_{int}^{N}$, 
\begin{equation}
H^{N}_{int}= p_{\rho}^{2}/m_{d} + 3p_{\lambda}^{2}/4m_{d} + V^{N}_{\textrm{conf}}(r_{\rho},r_{\lambda}),\label{Hint}
\end{equation}
where $V^{N}_{\textrm{conf}}=m_{d}\omega^{2}(r_{\rho}^{2}/4+r_{\lambda}^{2}/3)-3\omega/2$ is the aforementioned harmonic-oscillator potential. The terms of order $eB/m_{d}$ in eq.(\ref{kiny}) stem from the coupling of the conjugate momenta to $\tilde{\mathbf{A}}$, which are negligible at our order of approximation. Following refs.~\cite{IsgurKarl1978,IsgurKarl1979,Isgur}, approximate values for the constituent quark mass $m_{d}$ and the effective frequency $\omega$ are $m_{d}\approx340$MeV and $\omega\approx250$MeV, respectively. 

Owing to the approximate flavor symmetry of $H_{int}^{N}$ for $m_{d}\approx m_{u}$, we will see that $W$ does not affect the internal nucleon dynamics but for the recoil of the active quark.
At zero temperature, the internal nucleon state is the ground state of $H_{int}^{N}$, $|\phi_{int}^{0}\rangle$, whose energy we set to zero. In Jacobian spatial coordinates, its wavefunction is
\begin{equation}
\langle \mathbf{r}_{\lambda},\mathbf{r}_{\rho}|\phi_{int}^{0}\rangle=\frac{\beta^{3}}{3^{3/4}\pi^{3/2}}e^{-\beta^{2}(r_{\lambda}^{2}/3+r_{\rho}^{2}/4)},\quad\beta=\sqrt{\omega m_{d}}.\label{phinot}
\end{equation}
Excited states $|\phi_{int}^{m}\rangle$, $m\geq1$, become rapidly relativistic with energies $E_{int}^{m}=m\omega$. It is of note that, although $V^{N}_{\textrm{conf}}$ and $|\phi_{int}^{0}\rangle$ are manifestly model-dependent, they set generically an upper limit for the internal momenta of the constituent quarks at $\sim m_{d}$. Momentum conservation implies an equivalent cut-off for the momentum of the virtual leptons which interact with the active quark.

Finally, the initial state of the system is the ground state of $H_{0}$, made of an unpolarized proton at rest and the leptonic vacuum, at zero magnetic field. Using the nomenclature of the precedent section, it reads 
\begin{equation}
|\Phi(0)\rangle=|\Phi\rangle_{\mathbf{0}}\equiv|\Omega_{l}\rangle_{\mathbf{0}}\otimes|\phi_{int}^{0}\rangle\otimes[|0,1;\mathbf{0}\rangle+|0,-1;\mathbf{0}\rangle]^{p}_{\mathbf{0}}/2Lm_{p}^{1/2},\label{phi0}
\end{equation}
where $L\rightarrow\infty$ is a normalization length and the subscript $\mathbf{0}$ stands for zero magnetic field.
 
\subsection{Effective four-fermion EW-interaction, $W$} 

On top of $H_{0}$ the EW interaction, $W$, is perturbative. Since the upper limit for the energy of the intermediate states in the process of figure~\ref{fig1} is of the order of $m_{d}$, thus much less than the mass of W$^{\pm}$ bosons, 
$W$ is well described by the low-energy four-fermion potential \cite{FeynmanGellmann,Peskin,betadecay}, 
\begin{align}
&W=\frac{G_{F}}{\sqrt{2}}\int\textrm{d}^{3}R\:J_{h}^{\mu}(\tilde{\mathbf{r}})J^{l}_{\mu}(\tilde{\mathbf{r}})+\textrm{h.c.},\:\:\mu=0,1,2,3,\label{Wpotential}\\
&\textrm{where }\:\:J^{l}_{\mu}(\tilde{\mathbf{r}})=\bar{\Psi}_{e}(\tilde{\mathbf{r}})\gamma^{\mu}(\mathbb{I}-\gamma_{5})\Psi_{\nu}(\tilde{\mathbf{r}})\textrm{ and }J_{h}^{\mu}(\tilde{\mathbf{r}})=\bar{\Psi}_{p}(\tilde{\mathbf{r}})\gamma^{\mu}(\mathbb{I}-g_{A}\gamma_{5})\Psi_{n}(\tilde{\mathbf{r}})\nonumber
\end{align}
are the leptonic and hadronic currents respectively, with $g_{A}\simeq1.26$
being the axial-vector coefficient and $G_{F}$ the Fermi constant.  As mentioned earlier, the EW interaction takes place within the nucleon at the location of the active quark, $\tilde{\mathbf{r}}$, while the remaining quarks rest as spectators [figure~\ref{fig1}].   

Chirality in $W$ manifests in the coupling of left-handed particles to right-handed 
antiparticles. In turn, that results in a net alignment between the spins and momenta of the hadrons and the leptons which participate in the EW interaction. It explains, for instance, the celebrated experiment of Wu \emph{et al.} \cite{Wu} on the beta decay of 
polarized Co$^{60}$ nuclei, where left-handed electrons are preferably emitted in the direction opposite to the spins of the nuclei 
\cite{betadecay}. 

\section{Lepton Casimir momentum and proton kinetic energy}

In this section we compute the Casimir momentum of virtual leptons and explain the nature of the kinetic energy gained by an actual proton under the action of an adiabatically increasing magnetic field.

\subsection{Lepton Casimir momentum and proton velocity}\label{Adiabatic}

\begin{figure}[h]
\includegraphics[height=6.5cm,width=9.9cm,clip]{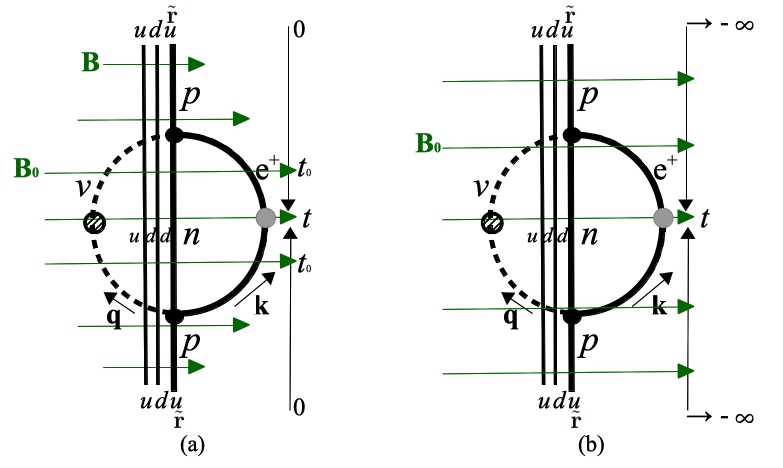}
\caption{(a) Diagrammatic representation of $\langle(\mathbf{K}_{e^{+}}+\mathbf{Q}_{\nu})(t)\rangle$ at order $W^{2}$ considering the time evolution of $\mathbf{B}$.  Time runs along the vertical. The magnetic field increases from $\mathbf{0}$ at time $0$ to $\mathbf{B}_{0}$ at $t_{0}$. %Lepton kinetic momentum operators apply at the observation time $t\gtrsim t_{0}$. 
 Black circles depict vertex operators, $W$. Kinetic momentum operators of neutrinos (dashed circle) and positrons (grey circle) apply at the observation time $t\gtrsim t_{0}$. %depict the kinetic momentum operators of neutrinos and positrons, respectively. 
 (b) Diagrammatic representation of $\langle\mathbf{K}_{e^{+}}+\mathbf{Q}_{\nu}\rangle_{\mathbf{B}_{0}}$  according to eqs.(\ref{equ}) and (\ref{equationmother}) in the adiabatic approximation. The value of the magnetic field remains constant at $\mathbf{B}_{0}$, and the asymptotic state in the far past is $|\Phi\rangle_{\mathbf{B}_{0}}$ given in eq.(\ref{phiBo}).}\label{fig1} 
 \end{figure}

We show by direct computation that the application of an adiabatically increasing magnetic field $\mathbf{B}$ on an initially 
unpolarised proton at rest, makes virtual leptons acquire a net linear momentum parallel to $\mathbf{B}$ while the nucleon accelerates in the opposite direction. The quantity of our interest is thus $\langle(\mathbf{K}_{e^{+}}+\mathbf{Q}_{\nu})(t)\rangle$, where $t$ is the time of observation and $\mathbf{K}_{e^{+}}$, $\mathbf{Q}_{\nu}$ are the kinetic momentum operators of positrons and neutrinos, respectively,
\begin{equation}
\mathbf{Q}_{\nu}=\int\textrm{d}^{3}r\:\Psi^{\dagger}_{\nu}(\mathbf{r})(-i\mathbf{\nabla})\Psi_{\nu}(\mathbf{r})|_{\textrm{neutrinos}},\quad
\mathbf{K}_{e^{+}}=\int\textrm{d}^{3}r\:[\Psi^{\dagger}_{e}(\mathbf{r})(-i\mathbf{\nabla})\Psi_{e}(\mathbf{r})+e\tilde{\mathbf{A}}]|_{\textrm{positrons}}.\nonumber
\end{equation}
The expressions for $\mathbf{K}_{e^{+}}$ and $\mathbf{Q}_{\nu}$ in terms of creation and annihilation operators of positrons, $b_{\mathcal{N},\mathcal{S}}^{e(\dagger)}(k_{x},k_{z})$, and neutrinos, $a^{\nu(\dagger)}_{\mathcal{S}}(\mathbf{q})$, are derived in appendix \ref{KQ}. The chirality of $W$ along with the magnetic field $\mathbf{B}$ break the mirror symmetry with respect to the $xy$-plane. Therefore, the only nonvanishing component of  $\langle(\mathbf{K}_{e^{+}}+\mathbf{Q}_{\nu})(t)\rangle$ may be that along $\mathbf{B}\parallel\hat{\mathbf{z}}$. In addition, translation invariance of $H$ along the magnetic field implies conservation of total momentum in that direction. Writing the kinetic momentum operator of the nucleon doublet as
\begin{equation}
\mathbf{P}_{N}=\int\textrm{d}^{3}r\Bigl[\Psi^{\dagger}_{n}(\mathbf{r})(-i\mathbf{\nabla})\Psi_{n}(\mathbf{r})|_{\textrm{neutrons}}
+\Psi^{\dagger}_{p}(\mathbf{r})(-i\mathbf{\nabla})\Psi_{p}(\mathbf{r})-e\tilde{\mathbf{A}}]|_{\textrm{protons}}\Bigr],\nonumber
\end{equation}
that implies $\hat{\mathbf{z}}\cdot[\mathbf{P}_{N}+\mathbf{K}_{e^{+}}+\mathbf{Q}_{\nu},\:H]=0$, from which it holds that $\hat{\mathbf{z}}\cdot\langle\mathbf{P}_{N}(t)\rangle=-\hat{\mathbf{z}}\cdot\langle(\mathbf{K}_{e^{+}}+\mathbf{Q}_{\nu})(t)\rangle$. Therefore, if the leptonic momentum presents a non-zero component along $\mathbf{B}$, we can be certain that the  nucleon accelerates in the opposite direction. Whereas the kinetic momentum of the actual nucleon, $\langle\mathbf{P}_{N}(t)\rangle$, is the quantity that can be accessed experimentally, the momentum of the virtual leptons, $\langle(\mathbf{K}_{e^{+}}+\mathbf{Q}_{\nu})(t)\rangle$, is what we can compute. In order to do so, we apply the formalism of appendix \ref{Dyna} to the expectation value of the leptonic momentum as the magnetic field increases adiabatically from $\mathbf{0}$ to its final value $\mathbf{B}_{0}$ at time $t_{0}$. Considering the observation time $t$ greater than $t_{0}$ and much shorter than the  spin-relaxation time of the proton, it holds that $\langle(\mathbf{K}_{e^{+}}+\mathbf{Q}_{\nu})(t)\rangle=\langle(\mathbf{K}_{e^{+}}+\mathbf{Q}_{\nu})(t_{0})\rangle$ for $t\gtrsim t_{0}$. Using the nomenclature of appendix \ref{Dyna} and eq.(\ref{Ototstat}), we can write this quantity in the adiabatic approximation as
\begin{equation}
\langle\mathbf{K}_{e^{+}}+\mathbf{Q}_{\nu}\rangle_{\mathbf{B}_{0}}=\:_{\mathbf{B}_{0}}\langle\Phi|\:\tilde{\mathbb{U}}^{\dagger}_{\mathbf{B_{0}}}(t)(\mathbf{K}_{e^{+}}+\mathbf{Q}_{\nu})\tilde{\mathbb{U}}_{\mathbf{B}_{0}}(t)\:|\Phi\rangle_{\mathbf{B}_{0}},\quad t\gtrsim t_{0},\label{equ}
\end{equation}
where $|\Phi\rangle_{\mathbf{B}_{0}}$ is the state of the system which evolves adiabatically from $|\Phi(0)\rangle$ towards its corresponding eigenstate of $H_{0}(\mathbf{B}_{0})$,
\begin{equation}
|\Phi\rangle_{\mathbf{B}_{0}}=|\Omega_{l}\rangle_{\mathbf{B}_{0}}\otimes|\phi_{int}^{0}\rangle\otimes\bigl[\bigl|0,1;\mathbf{0}\bigl\rangle+|0,-1;\mathbf{0}\bigl\rangle\bigr]^{p}_{\mathbf{B}_{0}}/2L\sqrt{E_{0}^{p}},\label{phiBo}
\end{equation}
and  $\tilde{\mathbb{U}}_{\mathbf{B}_{0}}(t)$ is given in eq.(\ref{eta}). We compute eq.(\ref{equ}) at leading order in $W$, i.e., $\mathcal{O}(W^{2})$.  Its diagrammatic representation 
 is that of figure~\ref{fig1}(b). There, an actual proton %, stable for $\mu_{N}B\ll m_{n}-m_{p}$, 
emits a virtual positron-neutrino pair in the presence of an external magnetic field $\mathbf{B}_{0}$.  Applying the aforementioned symmetry considerations to eq.(\ref{equ}), the Casimir momentum of leptons at time $t\gtrsim t_{0}$ is, at $\mathcal{O}(W^{2})$, %\cite{Peskin},
\begin{align}
&\langle\mathbf{K}_{e^{+}}+\mathbf{Q}_{\nu}\rangle_{\mathbf{B}_{0}}=\frac{G_{F}^{2}}{2}\hat{\mathbf{z}}\int\textrm{d}^{3}R\textrm{d}^{3}R'\:_{\mathbf{B}_{0}}\langle\Phi|J_{h}^{\mu}(\tilde{\mathbf{r}})J^{l}_{\mu}(\tilde{\mathbf{r}})\frac{(\mathbf{K}_{e^{+}}+\mathbf{Q}_{\nu})\cdot\hat{\mathbf{z}}}{[H_{0}(\mathbf{B}_{0})-E_{0}^{p}]^{2}}J^{l\dagger}_{\rho}(\tilde{\mathbf{r}}')J_{h}^{\rho\dagger}(\tilde{\mathbf{r}}')|\Phi\rangle_{\mathbf{B}_{0}},\nonumber\\
&\textrm{where}\quad\mu,\rho=0,..,3,\quad E_{0}^{p}=m_{p}+\mu_{N}B_{0},\quad\textrm{and}\label{equationmother}
\end{align}
\begin{equation}
(\mathbf{Q}_{\nu}+\mathbf{K}_{e^{+}})\cdot\hat{\mathbf{z}}=\int\frac{\textrm{d}^{3}q}{(2\pi)^{3}}\:q_{z}\sum_{\mathcal{S}=\pm1}a^{\nu\dagger}_{\mathcal{S}}(\mathbf{q})a^{\nu}_{\mathcal{S}}(\mathbf{q})+\int\frac{\textrm{d}k_{x}\textrm{d}k_{z}}{(2\pi)^{2}}k_{z}\sum_{\mathcal{S}=\pm1}\sum_{\mathcal{N}=0}b^{e\dagger}_{\mathcal{N},\mathcal{S}}(k_{x},k_{z})b^{e}_{\mathcal{N},\mathcal{S}}(k_{x},k_{z}).\nonumber
\end{equation}
The leptonic and hadronic currents flanking the energy denominator of eq.(\ref{equationmother}) give rise to quadratic vacuum fluctuations for leptons, and quadratic hadron fluctuations upon the proton state $\bigl[\bigl|0,1;\mathbf{0}\bigl\rangle+|0,-1;\mathbf{0}\bigl\rangle\bigr]^{p}_{\mathbf{B}_{0}}/2L\sqrt{E_{0}^{p}}$. Explicit expressions of these fluctuations are written in appendix \ref{AppC}. %It is of note that $\langle(\mathbf{K}_{e^{+}}+\mathbf{Q}_{\nu})(t)\rangle$ remains constant for $t\gtrsim t_{0}$ as long as the state $|\Phi\rangle_{\mathbf{B}_{0}}$ remains stationary too. 

In eq.(\ref{equationmother}), $\sum_{\alpha}H_{mag}^{\alpha}$ in $H_{0}(\mathbf{B}_{0})$ causes an  effective orientation of the spin of each magnetic particle along $\mathbf{B}_{0}$, $\alpha=e,n,p$. On the other hand, the chiral potential $W$ induces an alignment between the spins of the particles and their momenta. Both effects together result in a net alignment of the linear momenta of leptons and hadrons along $\mathbf{B}_{0}$ and, in turn, in a nonvanishing value of $\langle\mathbf{K}_{e^{+}}+\mathbf{Q}_{\nu}\rangle_{\mathbf{B}_{0}}$.  Its final expression reads 
%\begin{widetext}
\begin{subequations}\label{eq:fa}
\begin{align}
\label{eq:fa:1}
\langle\mathbf{K}_{e^{+}}&+\mathbf{Q}_{\nu}\rangle_{\mathbf{B}_{0}}\simeq\textrm{Re}\int\textrm{d}y\:\textrm{d}y'\int\frac{\textrm{d}p_{y}\textrm{d}^{3}q\textrm{d}k_{x}\textrm{d}k_{z}}{(2\pi)^{6}}%\int\frac{\textrm{d}^{3}q}{(2\pi)^{3}}\int\frac{\textrm{d}k_{x}\textrm{d}k_{z}}{(2\pi)^{2}}
 e^{i(p_{y}+q_{y})(y-y')}(k_{z}+q_{z})\:\hat{\mathbf{z}}\sum_{m=0}^{\infty}\Bigl|\langle\tilde{\phi}_{int}^{0}|\tilde{\phi}_{int}^{m}\rangle\Bigr|^{2}\nonumber\\
&\times\sum_{\mathcal{N}=0}^{\infty}\frac{G_{F}^{2}}{q\:\mathcal{E}_{\mathcal{N},m}^{2}E^{e}_{\mathcal{N}}}\Bigg\{I_{0}(\xi_{p})I_{0}(\xi^{'}_{p})(\mathbf{k}\cdot\langle\delta\mathbf{S}_{e^{+}}^{\mathcal{N}}\rangle)\bigg[q\:(1+3g^{2}_{A})+\frac{\mathbf{q}\cdot\mathbf{p}}{m_{n}}(g_{A}-1)^{2}\bigg]+I_{\mathcal{N}}(\xi_{e^{+}})\nonumber\\
%&\times(g_{A}-1)^{2}\Bigr]
&\times I_{\mathcal{N}}(\xi^{'}_{e^{+}})\bigg[ (\mathbf{k}\cdot\langle\delta\mathbf{S}_{p}\rangle)q\:(g_{A}-1)^{2}+ (\mathbf{q}\cdot\langle\delta\mathbf{S}_{p}\rangle)E^{e}_{\mathcal{N}}(g_{A}+1)^{2}\bigg]+I_{\mathcal{N}}(\xi_{e^{+}})I_{\mathcal{N}}(\xi^{'}_{e^{+}})\nonumber\\
&\times I_{0}(\xi_{p})I_{0}(\xi^{'}_{p})\bigg[(\mathbf{k}\cdot\langle\delta\mathbf{S}_{n}\rangle)\:2q\:(1+3g^{2}_{A}+4g_{A})+ (\mathbf{q}\cdot\langle\delta\mathbf{S}_{n}\rangle)E^{e}_{\mathcal{N}}(g^{2}_{A}-1)\bigg]\Bigg\},%\label{f1a}
\end{align}
\begin{align}
&\textrm{ where }\langle\delta\mathbf{S}_{e^{+}}^{\mathcal{N}}\rangle=\frac{e\mathbf{B}_{0}}{2(E^{e}_{\mathcal{N}})^{2}}I_{\mathcal{N}}(\xi_{e^{+}})I_{\mathcal{N}}(\xi^{'}_{e^{+}}),\:\:\langle\delta \mathbf{S}_{p}\rangle=\frac{\mu_{N}\mathbf{B}_{0}}{2m_{p}}[I_{0}(\xi_{p})I_{0}(\xi^{'}_{p})+I_{1}(\xi_{p})I_{1}(\xi^{'}_{p})],\nonumber\\
&\textrm{ and }\langle\delta \mathbf{S}_{n}\rangle=\frac{|\kappa_{n}|\mu_{N}\mathbf{B}_{0}}{4m_{n}}\label{eq:fa:2}%\label{spins}
\end{align}
\end{subequations}
%\end{widetext}
are, respectively, the net spin density of any pair of consecutive positron Landau levels, $|\mathcal{N},-1;\mathbf{k}\rangle^{e}_{\mathbf{B}_{0}}$ and $|\mathcal{N}+1,1;\mathbf{k}\rangle^{e}_{\mathbf{B}_{0}}$; the net spin density of the two lowest proton Landau levels, $|0,1;\mathbf{0}\rangle^{p}_{\mathbf{B}_{0}}$ and $|0,-1;\mathbf{0}\rangle^{p}_{\mathbf{B}_{0}}$; and the net spin of any pair of neutron states with equal momenta and different spin numbers, -1 and +1. Their equations are written in appendix \ref{spines}. In eq.(\ref{eq:fa:1}), $y$ and $y'$ are the coordinates of the center of mass of the nucleon along $\hat{\mathbf{y}}$; $\mathbf{k}$ is the canonical momentum of the virtual positrons; $\mathbf{q}$ is the momentum of the virtual neutrinos; and $\mathbf{p}$ is the momentum of the intermediate neutron. %At each vertex, total momentum fails to be conserved on the plane perpendicular to $\mathbf{B}_{0}$. 
 With our choice of gauge, conservation of momentum along the axis $\hat{\mathbf{x}}$ and $\hat{\mathbf{z}}$ holds at each vertex, implying $p_{x,z}=-(k_{x,z}+q_{x,z})$. Concerning the energy denominators, 
for $eB_{0}\ll m^{2}_{d}$, $\mathcal{E}_{\mathcal{N},m}\approx(k_{z}^{2}+q^{2}+2\mathcal{N}eB_{0})^{2}/2m_{d}+q_{z}k_{z}/m_{d}+q+E_{\mathcal{N}}^{e}+m_{n}-m_{p}+m\omega$ is the energy difference between the initial and intermediate states. The derivation of the kinetic terms is detailed in appendix \ref{AppD1}. The functions $I_{\mathcal{N}}(\xi)=(eB_{0})^{1/4}H_{\mathcal{N}}(\xi)e^{-\xi^{2}/2}/\pi^{1/4}\sqrt{2^{\mathcal{N}}\mathcal{N}!}$ are the charge distribution functions of the Landau levels, with $H_{\mathcal{N}}$ the Hermite polynomial of order $\mathcal{N}$, $\xi^{(')}_{p}=\sqrt{eB_{0}}y^{(')}$ for protons, and $\xi^{(')}_{e^{+}}=\sqrt{eB_{0}}y^{(')}+k_{x}/\sqrt{eB_{0}}$ for positrons. Finally, $|\langle\tilde{\phi}_{int}^{0}|\tilde{\phi}_{int}^{m}\rangle|^{2}$ is the square of the form factor between the internal nucleon ground state and the $m^{th}$ state,
\begin{equation}
\langle\tilde{\phi}_{int}^{0}|\tilde{\phi}_{int}^{m}\rangle=\langle\phi^{0}_{int}|e^{\pm2i[\mathbf{r}_{\lambda}\cdot(\mathbf{k}+\mathbf{q})+r_{\lambda}^{y}(\pm\sqrt{2\mathcal{N}eB}-k_{y})]/3}|\phi_{int}^{m}\rangle.\nonumber
\end{equation}
For the nonrelativistic harmonic oscillator in eq.(\ref{Hint}) the form factors posses a Gaussian profile in momentum space  which, in an effective manner, sets a cutoff for the momentum integrals at $\sim m_{d}$ --see eq.(\ref{formy}).
 
As anticipated, eq.(\ref{eq:fa:2}) shows that the effective spin-polarization of protons, positrons and neutrons is proportional to the magnetic field. In addition, the alignment of the momenta of virtual positrons and neutrinos with the spins reflects in the pseudo-scalar products of spins and momenta in the integrand of eq.(\ref{eq:fa:1}). As in Wu's experiment \cite{Wu,betadecay}, that causes virtual positrons and neutrinos to be emitted and absorbed with different probabilities in the directions parallel and antiparallel to $\mathbf{B}_{0}$. In our case, in virtue of total momentum conservation along $\mathbf{B}_{0}$, that results in a net transfer of momentum to the actual proton. %The dominant terms are those proportional to $\langle\delta\mathbf{S}_{e^{+}}^{\mathcal{N}}\rangle$. 
Passing the sums to continuum integrals and using the nonrelativistic harmonic-oscillator model of Isgur \& Karl for the internal eigenstates of the nucleon \cite{Coince,IsgurKarl1978,IsgurKarl1979}, we arrive at 
\begin{subequations}\label{eq:fb}
\begin{align}
\langle\mathbf{K}_{e^{+}}+\mathbf{Q}_{\nu}\rangle_{\mathbf{B}_{0}}
&\approx\frac{G_{F}^{2}m_{d}^{3}}{160\pi^{4}}\Big[(1+3g_{A}^{2})-\frac{4m_{d}}{m_{n}}(g_{A}-1)^{2}
%\nonumber\\&\qquad\qquad\quad\:\:
+\frac{2m^{2}_{d}}{m^{2}_{n}}(10g^{2}_{A}+3g_{A}+9)\Big]e\mathbf{B}_{0},\label{f1b}\\%,\:t\gtrsim t_{0}
\langle\mathbf{v}_{p}\rangle_{\mathbf{B}_{0}}&\approx-1.5\cdot10^{-15}\frac{e\mathbf{B}_{0}}{m_{d}m_{p}},\textrm{ in natural units with }\hbar=c=1,\label{f1b2}
\end{align}
\end{subequations}
for the leptonic Casimir momentum and the proton velocity, respectively. It is of note that, in analogy with the calculation of the Lamb-shift for the hydrogen atom, it is the inclusion of retardation factors in the lepton fields, together with the internal wavefunction of the nucleon, which renders the momentum integrals finite --cf. refs.~\cite{Milonnibook,Bethe} and appendix \ref{AppD}.  

\subsection{Proton kinetic energy, magnetic energy, and electroweak work}\label{AdiaEnergy}

In the first place, the kinetic energy of the proton, $m_{p}\langle\mathbf{v}_{p}\rangle_{\mathbf{B}_{0}}^{2}/2=\langle\mathbf{P}_{N}\rangle_{\mathbf{B}_{0}}^{2}/2m_{p}$, is a  
Doppler-shift correction to its EW self-energy. The proof of this statement involves a number of technical issues which are  detailed in appendix \ref{AppE}. There, $\langle W\rangle_{\mathbf{B}_{0}}$ is computed integrating the variations of the EW energy, $\delta\langle W\rangle_{\mathbf{B}}$, from $\mathbf{B}=\mathbf{0}$ to $\mathbf{B}=\mathbf{B}_{0}$. It is shown that $\delta\langle W\rangle_{\mathbf{B}}$ contains a term 
\begin{equation}
-\langle\mathbf{P}_{N}\rangle_{\mathbf{B}}\cdot\delta\langle\mathbf{K}_{e^{+}}+\mathbf{Q}_{\nu}\rangle_{\mathbf{B}}/m_{p},\label{Doppler1}
\end{equation}
which is readily identifiable with the energy associated to the differential Doppler-shift experienced  by the virtual leptons emitted and absorbed by the proton in motion along $\mathbf{B}$. The integration of eq.(\ref{Doppler1}) leads straightaway to $\langle\mathbf{P}_{N}\rangle_{\mathbf{B}_{0}}^{2}/2m_{p}$.

Next, the question arises of what is the source of this energy.  It is immediate to prove that (see appendix \ref{AppE}) 
\begin{equation}
\frac{\delta\langle W\rangle_{\mathbf{B}}}{\delta\mathbf{B}}\quad\textrm{is part of}\quad
\left\langle\frac{\delta H_{mag}^{p}+\delta H_{mag}^{e}+\delta H_{mag}^{n}}{\delta\mathbf{B}}\right\rangle_{\mathbf{B}},\nonumber
\end{equation}
and hence, so are the variations of the kinetic energy $\frac{\delta\langle\mathbf{P}_{N}\rangle_{\mathbf{B}}^{2}}{2m_{p}\delta\mathbf{B}}$. Therefore we conclude that, while the kinetic energy of the proton is a magnetic energy supplied by the source of magnetic field, its kinetic momentum along $\mathbf{B}$ is provided by the virtual leptons which perform an electroweak work upon the proton. The magnetic field does not perform any work by itself. On physical grounds, the magnetic energy is spent in breaking the P and T symmetries in the system, which in turn causes the asymmetry on the linear momentum of the virtual leptons which propagate in the directions parallel and antiparallel to $\mathbf{B}.$

\section{Conclusions and Discussion}\label{Conclusions}

We have shown that a single proton accelerates along an adiabatically increasing magnetic field, $\mathbf{B}$, as a result of the EW self-interaction of the active quark within the nucleon. Reciprocally, the virtual leptons which participate in the interaction acquire a Casimir momentum in the opposite direction. At leading order, that momentum is proportional  to the magnetic field and to the square of Fermi's constant [eqs.(\ref{f1b}) and (\ref{f1b2})]. Regarding energetics, the proton kinetic energy is a magnetic-induced Doppler-shift correction to its EW self-energy.  It follows that, while the kinetic momentum of the proton  is transferred from the virtual leptons, its kinetic energy is supplied by the source of the magnetic field. We interpret that the magnetic energy is spent in breaking the P and T symmetries of the system along $\mathbf{B}$. In turn, that causes an asymmetry on the variation of the momentum of the virtual leptons which propagate parallel and antiparallel to $\mathbf{B}$. The net reaction force of those leptons on the proton does the electroweak work which accelerates it. 
 
In physical grounds, our findings in EW and those of ref.~\cite{PRLDonaire} in QED imply that, generically, a magnetic particle accelerates linearly along a spatially uniform magnetic field if its interaction with the quantum vacuum is chiral. Reciprocally, the quantum vacuum fluctuations gain a Casimir momentum. Ultimately, since the interaction of any material particle with the EW vacuum is chiral, it follows that, in the presence of a magnetic field, the EW Casimir momentum is a generic phenomenon in nature.

Regarding the approach, some comments are in order. The results of eqs.(\ref{f1b}) and (\ref{f1b2}) depend ultimately on the hadron model which determines the nucleon wavefunction in eq.(\ref{eq:fa:1})  \cite{Coince,IsgurKarl1978,IsgurKarl1979,deRujula,Isgur}. Thus, although our simplified harmonic oscillator model is expected to yield the correct orders of magnitude on the estimates, corrections of order unity are to be contemplated. Nonetheless, this makes the leptonic Casimir momentum a suitable phenomenon to test hadron models at low energies \cite{Shulga,cavitychromo,QuarkModels}.  For the values of laboratory magnetic fields, our estimate for the proton speed results extremely small. However, for the strong magnetic fields of magnetars, $B\sim10^{14}$T \cite{Khalilov2005,Mao}, we obtain $v_{p}\sim10\:$nm/s.

Concerning the physical interpretation of our findings, the Casimir momentum is to be attributed to virtual particles  coupled to actual particles, and not to the quantum vacuum itself. The latter interpretation seems to rest on the erroneous assumption that the vacuum state  and the state of the actual particles are disentangled --cf.  ref.~\cite{JPCMDonaire}. In our case, the fact that the Casimir momentum is carried by virtual leptons, thus being physically inaccessible, does not mean that the Casimir momentum is carried off by the EW vacuum alone.  Conversely, the fact that only the kinetic momentum of the actual proton is physically accessible does not imply violation of momentum conservation. Hence, it is the total momentum of the nucleon-EW-vacuum state considered as a whole which remains null, in accordance with momentum conservation and global translation invariance along $\mathbf{B}$.

Finally, let us outline some perspectives for future work. It is plain from eqs.(\ref{eq:fa:1}) and (\ref{eq:fa:2}) that the dependence of the Casimir momentum on $\mathbf{B}$ is due to the implicit dependence of the particle spins on the magnetic field. Thus, eq.(\ref{eq:fa:1}) suggests that, in our scenario and at leading order in $W$, the maximum value of the Casimir momentum is achieved for a fully-polarized proton. We plan to apply the present formalism to the calculation of the Casimir momentum and the acceleration of a proton as it gets spin-polarized,  either through spontaneous spin-relaxation \cite{spinrelax} or through nuclear magnetic resonance \cite{NMR,PRL106_253001(2011)}.\\

\noindent\textbf{Acknowledgments}\\

\noindent We thank A. Cano, L. Froufe, M. Chernodub, G. Rikken and B. van Tiggelen for fruitful discussions on the subject.
Financial support from grants MTM2014-57129-C2-1-P (MINECO), BU229P18 and VA137G18 (JCyL) is acknowledged.\\

\appendix

\section{Spinor eigenstates of magnetic leptons and nucleons in a uniform and stationary magnetic field}\label{AppA}

In this appendix we compile the equations for the spinor eigenstates of the Hamiltonian $H_{D}^{\alpha}+H_{mag}^{\alpha}$ for electrons ($\alpha=e$), protons ($\alpha=p$) and neutrons ($\alpha=n$) in the presence of an external and uniform magnetic field, $\mathbf{B}=B\hat{\mathbf{z}}$, in the gauge $\tilde{\mathbf{A}}=-Br_{y}\hat{\mathbf{x}}$. We make use of the results within refs.~\cite{Mao,Bhattacharya,Bander,Broderick}, as well as the textbooks of Peskin \& Schroeder and Itzykson \& Zube \cite{Peskin}. We employ the chiral representation, with the Dirac matrices given by
\[
  \gamma_{0}=
  \left[ {\begin{array}{cc}
  0 & - \mathbb{I}\\       -\mathbb{I}&0\      \end{array} } \right],\quad \gamma_{5}=
  \left[ {\begin{array}{cc}
    \mathbb{I} & 0\\       0 & -\mathbb{I}\      \end{array} } \right],\quad \gamma_{i}=
  \left[ {\begin{array}{cc}
    0 & \sigma_{i}\\       -\sigma_{i} &0\      \end{array} } \right],
\]
where $\mathbb{I}$  is the two-dimensional identity matrix and $\sigma_{i}$ is the $i^{th}$ Pauli matrix. In this representation, $\gamma_{5}$ writes $\gamma_{5}=+i\gamma_{0}\gamma_{1}\gamma_{2}\gamma_{3}$.

\subsection{Electron and positron Landau levels}\label{posit}

For electrons, the eigenstates of the Hamiltonian density operator, $\hat{h}_{D}^{e}+\hat{h}_{mag}^{e}=\gamma_{0}[-i\gamma^{j}\partial_{j}-eB\:r_{y}\gamma_{1}+m_{e}\mathbb{I}]$, $j=1,2,3$, correspond to the spinors of Landau levels with positive and negative energies, $E^{e}_{\mathcal{N}}=\sqrt{m_{e}^{2}+k_{z}^{2}+2\mathcal{N}eB}$, where $\mathcal{N}$ is the principal quantum number, $\mathcal{N}=0,1,2,3,...$, and $k_{z}$ is the momentum component along $\mathbf{B}$. Both electron and positron eigenstates are degenerate with respect to the spin number, $\mathcal{S}=\pm1$, except for the ground state $\mathcal{N}=0$ for which a unique solution exists with $\mathcal{S}=-1$. 
Denoting $\xi_{e}=\sqrt{eB}(r_{y}-\frac{k_{x}}{eB})$ and $\xi_{e^{+}}=\sqrt{eB}(r_{y}+\frac{k_{x}}{eB})$, the spinor eigenstates for electrons read
\[
U^{e}_{\mathcal{N},+}=\left[ \begin{array}{c} \frac{E^{e}_{\mathcal{N}}+m_{e}+k_{z}}{\sqrt{2(E^{e}_{\mathcal{N}}+m_{e})}}I_{\mathcal{N}-1}(\xi_{e})\\\\-\sqrt{\frac{\mathcal{N}eB}{E^{e}_{\mathcal{N}}+m_{e}}}I_{\mathcal{N}}(\xi_{e})\\\\-\frac{E^{e}_{\mathcal{N}}+m_{e}-k_{z}}{\sqrt{2(E^{e}_{\mathcal{N}}+m_{e})}}I_{\mathcal{N}-1}(\xi_{e})\\\\-\sqrt{\frac{\mathcal{N}eB}{E^{e}_{\mathcal{N}}+m_{e}}}I_{\mathcal{N}}(\xi_{e})\end{array} \right],\quad
  U^{e}_{\mathcal{N},-}=\left[ \begin{array}{c}-\sqrt{\frac{\mathcal{N}eB}{E^{e}_{\mathcal{N}}+m_{e}}}I_{\mathcal{N}-1}(\xi_{e})\\\\\frac{E^{e}_{\mathcal{N}}+m_{e}-k_{z}}{\sqrt{2(E^{e}_{\mathcal{N}}+m_{e})}}I_{\mathcal{N}}(\xi_{e})\\\\-\sqrt{\frac{\mathcal{N}eB}{E^{e}_{\mathcal{N}}+m_{e}}}I_{\mathcal{N}-1}(\xi_{e})\\\\-\frac{E^{e}_{\mathcal{N}}+m_{e}+k_{z}}{\sqrt{2(E^{e}_{\mathcal{N}}+m_{e})}}I_{\mathcal{N}}(\xi_{e})\end{array} \right], \textrm{ for }\mathcal{S}=\pm1\textrm{ respectively;}
\]
while those for positrons are   
\[
V^{e}_{\mathcal{N},+}=\left[ \begin{array}{c}\frac{E^{e}_{\mathcal{N}}+m_{e}+k_{z}}{\sqrt{2(E^{e}_{\mathcal{N}}+m_{e})}}I_{\mathcal{N}-1}(\xi_{e^{+}})\\\\\sqrt{\frac{\mathcal{N}eB}{E^{e}_{\mathcal{N}}+m_{e}}}I_{\mathcal{N}}(\xi_{e^{+}})\\\\\frac{E^{e}_{\mathcal{N}}+m_{e}-k_{z}}{\sqrt{2(E^{e}_{\mathcal{N}}+m_{e})}}I_{\mathcal{N}-1}(\xi_{e^{+}})\\\\-\sqrt{\frac{\mathcal{N}eB}{E^{e}_{\mathcal{N}}+m_{e}}}I_{\mathcal{N}}(\xi_{e^{+}})\end{array} \right],\quad
  V^{e}_{\mathcal{N},-}=\left[ \begin{array}{c} \sqrt{\frac{\mathcal{N}eB}{E^{e}_{\mathcal{N}}+m_{e}}}I_{\mathcal{N}-1}(\xi_{e^{+}})\\\\\frac{E^{e}_{\mathcal{N}}+m_{e}-k_{z}}{\sqrt{2(E^{e}_{\mathcal{N}}+m_{e})}}I_{\mathcal{N}}(\xi_{e^{+}})\\\\-\sqrt{\frac{\mathcal{N}eB}{E^{e}_{\mathcal{N}}+m_{e}}}I_{\mathcal{N}-1}(\xi_{e^{+}})\\\\\frac{E^{e}_{\mathcal{N}}+m_{e}+k_{z}}{\sqrt{2(E^{e}_{\mathcal{N}}+m_{e})}}I_{\mathcal{N}}(\xi_{e^{+}})\end{array} \right], \textrm{ for }\mathcal{S}=\pm1\textrm{ respectively;}
\]
such that $\hat{h}_{D}^{e}e^{ik_{x}r_{x}+ik_{z}r_{z}}U^{e}_{\mathcal{N},\pm}=E^{e}_{\mathcal{N}}e^{ik_{x}r_{x}+ik_{z}r_{z}}U^{e}_{\mathcal{N},\pm}$, $\hat{h}_{D}^{e}e^{-ik_{x}r_{x}-ik_{z}r_{z}}V^{e}_{\mathcal{N},\pm}=-E^{e}_{\mathcal{N}}e^{-ik_{x}r_{x}-ik_{z}r_{z}}V^{e}_{\mathcal{N},\pm}$.    The functions $I_{\mathcal{N}}(X)$ were defined in section \ref{Adiabatic} as $I_{\mathcal{N}}(X)=(eB)^{1/4}H_{\mathcal{N}}(X)e^{-X^{2}/2}/\pi^{1/4}\sqrt{2^{\mathcal{N}}\mathcal{N}!}$, where $H_{\mathcal{N}}$ is the Hermite polynomial of order $\mathcal{N}$. In terms of the above eigenstates, the Dirac field of electrons reads, in Schr\"odinger's representation,
\begin{align}
\Psi_{e}(\mathbf{r})=\sum_{\mathcal{S}=\pm1}\sum_{\mathcal{N}=0}^{\infty}\int\frac{\textrm{d}k_{x}\textrm{d}k_{z}}{(2\pi)^{2}\sqrt{2E^{e}_{\mathcal{N}}}}&\bigl[a^{e}_{\mathcal{N},\mathcal{S}}(k_{x},k_{z})\:U^{e}_{\mathcal{N},\mathcal{S}}\:e^{i(k_{x}r_{x}+k_{z}r_{z})}\nonumber\\
&+b^{e\dagger}_{\mathcal{N},\mathcal{S}}(k_{x},k_{z})\:V^{e}_{\mathcal{N},\mathcal{S}}\:e^{-i(k_{x}r_{x}+k_{z}r_{z})}\bigr],\label{SM1}
\end{align}
where $a^{e(\dagger)}_{\mathcal{N},\mathcal{S}}(k_{x},k_{z})$ and $b^{e(\dagger)}_{\mathcal{N},\mathcal{S}}(k_{x},k_{z})$ are the annihilation and creation ($\dagger$) operators of electrons and positrons, respectively, which satisfy the usual anticommutation relations,
\begin{align}
\{a^{e}_{\mathcal{N},\mathcal{S}}(k_{x},k_{z}),a^{e\dagger}_{\mathcal{N}',\mathcal{S}'}(k_{x}',k_{z}')\}&=\{b^{e}_{\mathcal{N},\mathcal{S}}(k_{x},k_{z}),b^{e\dagger}_{\mathcal{N}',\mathcal{S}'}(k_{x}',k_{z}')\}\nonumber\\
&=(2\pi)^{2}\delta(k_{x}-k_{x}')\delta(k_{z}-k_{z}')\delta_{\mathcal{N}\mathcal{N}'}\delta_{\mathcal{S}\mathcal{S}'}.\label{SM2}
\end{align}
Positron states appear as vacuum fluctuations in the calculations of the Letter. In terms of positron creation operators acting on the leptonic vacuum, their expression is
\begin{equation}
|\mathcal{N},\mathcal{S};k_{x},k_{z}\rangle_{\mathbf{B}}^{e}=\sqrt{2E_{\mathcal{N}}^{e}}\:b_{\mathcal{N},\mathcal{S}}^{e\dagger}(k_{x},k_{z})|\Omega_{l}\rangle_{\mathbf{B}}.\nonumber
\end{equation}

\subsection{Low-lying proton Landau levels}

As for protons, we will restrict ourselves to the low-lying positive energy eigenstates of the Hamiltonian operator $\hat{h}_{D}^{p}+\hat{h}_{mag}^{p}=\gamma_{0}\bigl[-i\gamma^{j}\partial_{j}+eB\:r_{y}\gamma_{1}+m_{p}\mathbb{I}+i\kappa_{p}\mu_{N}B[\gamma_{2},\gamma_{1}]/4\bigr]$, $j=1,2$. These are, the positive energy spinors corresponding to the Landau levels with kinetic momentum  $p_{z}$ along $\mathbf{B}$, $|p_{z}|\ll\sqrt{eB}\ll m_{p}$, principal number $\mathcal{N}=0$ and spin numbers $S=+1$ and $S=-1$, with energies $E^{p}_{+}\simeq m_{p}+p_{z}^{2}/2m_{p}-\kappa_{p}eB/4m_{p}$ and $E^{p}_{-}\simeq m_{p}+p_{z}^{2}/2m_{p}+\kappa_{p}eB/4m_{p}+eB/m_{p}$, respectively,
\[
U^{p}_{0,+}\simeq i\sqrt{E^{p}_{+}}I_{0}(\xi_{p})\left[ \begin{array}{c}1\\\\0\\\\\ -1\\\\0\end{array} \right],\quad
  U^{p}_{0,-}\simeq (1-\frac{eB}{4m_{p}^{2}})\sqrt{E^{p}_{-}}\left[ \begin{array}{c}-i\frac{\sqrt{eB}}{\sqrt{2}m_{p}}I_{1}(\xi_{p})\\\\-I_{0}(\xi_{p})\\\\-i\frac{\sqrt{eB}}{\sqrt{2}m_{p}}I_{1}(\xi_{p})\\\\I_{0}(\xi_{p})\end{array} \right], \textrm{ for }\mathcal{S}=\pm1,\textrm{  respectively,}
\]
with $\xi_{p}=\sqrt{eB}(r_{y}+\frac{p_{x}}{eB})$, such that $(\hat{h}_{D}^{p}+\hat{h}_{mag}^{p})\:e^{i(p_{x}r_{x}+p_{z}r_{z})}U^{p}_{0,\pm}=E^{p}_{\pm}e^{i(p_{x}r_{x}+p_{z}r_{z})}U^{p}_{0,\pm}$.  
Creation and annihilation proton (and antiproton) operators follow anticommutation relations analogous to those in eq.(\ref{SM2}),
\begin{align}
\{a^{p}_{\mathcal{N},\mathcal{S}}(p_{x},p_{z}),a^{p\dagger}_{\mathcal{N}',\mathcal{S}'}(p_{x}',p_{z}')\}&=\{b^{p}_{\mathcal{N},\mathcal{S}}(p_{x},p_{z}),b^{p\dagger}_{\mathcal{N}',\mathcal{S}'}(p_{x}',p_{z}')\}\nonumber\\
&=(2\pi)^{2}\delta(p_{x}-p_{x}')\delta(p_{z}-p_{z}')\delta_{\mathcal{N}\mathcal{N}'}\delta_{\mathcal{S}\mathcal{S}'},\nonumber
\end{align}
and the proton field reads
\begin{align}
\Psi_{p}(\mathbf{r})=\sum_{\mathcal{S}=\pm1}\sum_{\mathcal{N}=0}^{\infty}\int\frac{\textrm{d}p_{x}\textrm{d}p_{z}}{(2\pi)^{2}\sqrt{2E^{p}_{\mathcal{N},\mathcal{S}}}}&\bigl[a^{p}_{\mathcal{N},\mathcal{S}}(p_{x},p_{z})\:U^{p}_{\mathcal{N},\mathcal{S}}\:e^{i(p_{x}r_{x}+p_{z}r_{z})}\nonumber\\&
+b^{p\dagger}_{\mathcal{N},\mathcal{S}}(p_{x},p_{z})\:V^{p}_{\mathcal{N},\mathcal{S}}\:e^{-i(p_{x}r_{x}+p_{z}r_{z})}\bigr],\nonumber
\end{align}
where $V^{p}_{\mathcal{N},\mathcal{S}}$ is the antiproton eigenstate and $E^{p}_{\mathcal{N},\mathcal{S}}$ the proton energy of the Landau level $\mathcal{N}$ with spin number $\mathcal{S}$. Relevant to our calculations are only the expressions for $U^{p}_{0,+}$ and $U^{p}_{0,-}$ given above.  
In terms of proton creation operators acting on the hadronic vacuum, $|\Omega_{h}\rangle_{\mathbf{B}}$, the two minimum energy Landau levels in the state of an unpolarized proton with transverse momentum $\mathbf{p}$ read 
\begin{equation}
|0,1;\mathbf{p}\rangle^{p}_{\mathbf{B}}=\sqrt{2E_{+}^{p}}a^{p\dagger}_{0,1}(p_{x},p_{z})|\Omega_{h}\rangle_{\mathbf{B}},\quad|0,-1;\mathbf{p}\rangle^{p}_{\mathbf{B}}=\sqrt{2E_{-}^{p}}a^{p\dagger}_{0,-1}(p_{x},p_{z})|\Omega_{h}\rangle_{\mathbf{B}}.\nonumber
\end{equation}

\subsection{Nonrelativistic neutron states and massless neutrino states}

Neutrons in the presence of a magnetic field $\mathbf{B}$ experience Zeeman splitting as a result of their anomalous magnetic moment. The eigenvalues of $\hat{h}_{D}^{n}+\hat{h}_{mag}^{n}=\gamma_{0}\bigl[-i\gamma^{j}\partial_{j}+m_{n}\mathbb{I}+i\kappa_{n}\mu_{N}B[\gamma_{2},\gamma_{1}]/4\bigr]$, $j=1,2,3$, for nonrelativistic neutrons of kinetic momentum $p\ll m_{n}$, and $eB\ll m_{n}^{2}$, depend on the quantum spin number, $\mathcal{S}=\pm1$, as $E^{n}_{\mathcal{S}}\simeq m_{n} + p^{2}/2m_{n}+\mathcal{S}|\kappa_{n}|\mu_{N}B/2$. In the  nonrelativistic limit, at leading order in $eB/m_{n}^{2}$, the neutron eigenstates are

\[
U^{n}_{\mathcal{S}}=\sqrt{E_{\mathcal{S}}^{n}}\frac{(1+p_{z}/2m_{n}-\mathcal{S}p_{z}|g_{n}|\mu_{N}B/4m^{2}_{n})(p_{x}+ip_{y})}{(\mathcal{S}+1)m_{n}+p_{x}+ip_{y}}\left[ \begin{array}{c}\frac{(\mathcal{S}+1)m_{n}+\mathcal{S}|p_{x}+ip_{y}|^2/2m_{n}}{p_{x}+ip_{y}}\\\\\mathcal{S}(1-p_{z}/m_{n})\\\\-\mathcal{S}(1-p_{z}/m_{n})\frac{(\mathcal{S}+1)m_{n}+\mathcal{S}|p_{x}+ip_{y}|^2/2m_{n}}{p_{x}+ip_{y}}\\\\1\end{array} \right]
\]
\[
+\sqrt{E_{\mathcal{S}}^{n}}\frac{(1+p_{z}/2m_{n})(p_{x}+ip_{y})|g_{n}|\mu_{N}B}{2(\mathcal{S}+1)m_{n}+p_{x}+ip_{y}}\left[ \begin{array}{c}0\\\\p_{z}/m_{n}^{2}\\\\-(\mathcal{S}+1)\frac{1+p_{z}/m_{n}}{p_{x}+ip_{y}}\\\\0\end{array} \right], \textrm{ for }\mathcal{S}=\pm1,
\] 
such that $(\hat{h}_{D}^{n}+\hat{h}_{mag}^{n})e^{i\mathbf{p}\cdot\mathbf{r}}U^{n}_{\mathcal{S}}=E_{\mathcal{S}}^{n}e^{i\mathbf{p}\cdot\mathbf{r}}U^{n}_{\mathcal{S}}$. The neutron field $\Psi_{n}(\mathbf{r})$ reads 
\begin{equation}
\Psi_{n}(\mathbf{r})=\sum_{\mathcal{S}=\pm1}\int\frac{\textrm{d}^{3}p}{(2\pi)^{3}}\frac{1}{\sqrt{2E^{n}_{\mathcal{S}}}}\bigl[a^{n}_{\mathcal{S}}(\mathbf{p})\:U^{n}_{\mathbf{p},\mathcal{S}}\:e^{i\mathbf{p}\cdot\mathbf{r}}+b^{n\dagger}_{\mathcal{S}}(\mathbf{p})\:V^{n}_{\mathbf{p},\mathcal{S}}\:e^{-i\mathbf{p}\cdot\mathbf{r}}\bigr],\nonumber
\end{equation}
where $V^{n}_{\mathbf{p},\mathcal{S}}$ is the antineutron eigenstate of spin $\mathcal{S}$ and momentum $\mathbf{p}$, and the creation/annihilation operators satisfy the usual anticommutation relations,
\begin{equation}
\{a^{n}_{\mathcal{S}}(\mathbf{p}),a^{n\dagger}_{\mathcal{S}'}(\mathbf{p}')\}=\{b^{n}_{\mathcal{S}}(\mathbf{p}),b^{n\dagger}_{\mathcal{S}'}(\mathbf{p}')\}=(2\pi)^{3}\delta^{(3)}(\mathbf{p}-\mathbf{p}')\delta_{\mathcal{S}\mathcal{S}'}.\nonumber
\end{equation}
In terms of neutron creation operators acting on the hadronic vacuum, $|\Omega_{h}\rangle_{\mathbf{B}}$, nonrelativistic neutron states read $|\mathbf{p};\mathcal{S}\rangle^{n}_{\mathbf{B}}=\sqrt{2E_{\mathcal{S}}^{n}}\:a^{n\dagger}_{\mathcal{S}}(\mathbf{p})|\Omega_{h}\rangle_{\mathbf{B}}.$\\\\

Lastly,  the  eigenstates of  neutrinos, which are ordinary free massless spinors, are
\[
U^{\nu}_{\mathcal{S}}=\sqrt{(E_{q}+q_{z})/2}\left[ \begin{array}{c}\mathcal{S}\frac{\sqrt{E_{q}^{2}-q_{z}^{2}}}{q_{x}+iq_{y}}\\\\\mathcal{S}\frac{\sqrt{E_{q}-q_{z}}}{\sqrt{E_{q}+q_{z}}}\\\\\frac{q_{z}-E_{q}}{q_{x}+iq_{y}}\\\\1\end{array} \right], \textrm{ for }\mathcal{S}=\pm1,\:\:E_{q}=q.
\]
The neutrino field $\Psi_{\nu}(\mathbf{r})$ reads 
\begin{equation}
\Psi_{\nu}(\mathbf{r})=\sum_{\mathcal{S}=\pm1}\int\frac{\textrm{d}^{3}q}{(2\pi)^{3}}\frac{1}{\sqrt{2E_{q}}}\bigl[a^{\nu}_{\mathcal{S}}(\mathbf{q})\:U^{\nu}_{\mathbf{q},\mathcal{S}}\:e^{i\mathbf{q}\cdot\mathbf{r}}+b^{\nu\dagger}_{\mathcal{S}}(\mathbf{q})\:V^{\nu}_{\mathbf{q},\mathcal{S}}\:e^{-i\mathbf{q}\cdot\mathbf{r}}\bigr],\nonumber
\end{equation}
where $V^{\nu}_{\mathbf{q},\mathcal{S}}$ is the antineutrino eigenstate of spin $\mathcal{S}$ and momentum $\mathbf{q}$, and the creation/annihilation operators satisfy the anticommutation relations,
\begin{equation}
\{a^{\nu}_{\mathcal{S}}(\mathbf{q}),a^{\nu\dagger}_{\mathcal{S}'}(\mathbf{q}')\}=\{b^{\nu}_{\mathcal{S}}(\mathbf{q}),b^{n\dagger}_{\mathcal{S}'}(\mathbf{q}')\}=(2\pi)^{3}\delta^{(3)}(\mathbf{q}-\mathbf{q}')\delta_{\mathcal{S}\mathcal{S}'}.\nonumber
\end{equation}
In terms of neutrino creation operators acting on the leptonic vacuum, $|\Omega_{l}\rangle_{\mathbf{B}}$, neutrino states read $|\mathbf{q};\mathcal{S}\rangle^{\nu}_{\mathbf{B}}=\sqrt{2E_{q}}\:a^{\nu\dagger}_{\mathcal{S}}(\mathbf{q})|\Omega_{l}\rangle_{\mathbf{B}}.$

\section{Spin expectation values and lepton kinetic momentum operators}\label{AppB}

We use the expressions for the spinor eigenstates of the electroweak particles to compute the spin expectation values which appear in eq.(\ref{eq:fa:2}) as well as the lepton kinetic operators, $\mathbf{K}_{e^{+}}$ and $\mathbf{Q}_{\nu}$.

\subsection{Spin expectation values}\label{spines}

Using the equations for the spinor eigenstates in appendix \ref{AppA}, we write down the expressions for the expectation values of the spin operators given in eq.(\ref{eq:fa:2}),
\begin{equation}
\langle\delta\mathbf{S}_{e^{+}}^{\mathcal{N}}\rangle\equiv\frac{\hat{\mathbf{z}}}{2E^{e}_{\mathcal{N}}}V_{\mathcal{N},+}^{e\dagger}S_{z}^{e^{+}}V_{\mathcal{N},+}^{e}+\frac{\hat{\mathbf{z}}}{2E^{e}_{\mathcal{N}-1}}V_{\mathcal{N}-1,-}^{e\dagger}S_{z}^{e^{+}}V_{\mathcal{N}-1,-}^{e}=\frac{e\mathbf{B}}{2(E^{e}_{\mathcal{N}})^{2}}I_{\mathcal{N}}(\xi_{e^{+}})I_{\mathcal{N}}(\xi^{'}_{e^{+}}),\nonumber
\end{equation}
\begin{equation}
\langle\delta\mathbf{S}_{p}\rangle\equiv\frac{\hat{\mathbf{z}}}{2E^{p}_{+}}U_{0,+}^{p\dagger}S_{z}^{p}\:U_{0,+}^{p}+\frac{\hat{\mathbf{z}}}{2E^{p}_{-}}U_{0,-}^{p\dagger}S_{z}^{p}\:U_{0,-}^{p}=\frac{\mu_{N}\mathbf{B}}{2m_{p}}[I_{0}(\xi_{p})I_{0}(\xi^{'}_{p})+I_{1}(\xi_{p})I_{1}(\xi^{'}_{p})],\nonumber
\end{equation}
\begin{equation}
\langle\delta\mathbf{S}_{n}\rangle\equiv\frac{\hat{\mathbf{z}}}{2E^{n}_{+}}U_{+}^{n\dagger}S_{z}^{n}\:U_{+}^{n}+\frac{\hat{\mathbf{z}}}{2E^{n}_{-}}U_{-}^{n\dagger}S_{z}^{n}\:U_{-}^{n}=\frac{|\kappa_{n}|\mu_{N}\mathbf{B}}{4m_{n}},\nonumber
\end{equation}
where $S_{z}^{e^{+}}$,  $S_{z}^{p}$ and  $S_{z}^{n}$ are the components of the spin operators along $\mathbf{B}$, for positrons, protons and neutrons, respectively.

\subsection{Kinetic momentum operators of leptons}\label{KQ}

In this subsection we write down the expressions of the kinetic momentum operators of positrons and neutrinos, $\mathbf{K}_{e^{+}}$ and $\mathbf{Q}_{\nu}$, in terms of their creation and annihilation operators. 

As for the neutrino momentum, it is,
\begin{equation}
\mathbf{Q}_{\nu}=\int\textrm{d}^{3}r\:\Psi^{\dagger}_{\nu}(\mathbf{r})(-i\mathbf{\nabla})\Psi_{\nu}(\mathbf{r})|_{\textrm{neutrinos}}=\int\frac{\textrm{d}^{3}q}{(2\pi)^{3}}\mathbf{q}\sum_{\mathcal{S}=\pm1}a^{\nu\dagger}_{\mathcal{S}}(\mathbf{q})a^{\nu}_{\mathcal{S}}(\mathbf{q}).\label{Qn}
\end{equation}

As for the kinetic momentum of positrons in a magnetic field $\mathbf{B}$, we make use of the formulas developed in appendix \ref{posit} in the gauge $\tilde{\mathbf{A}}=-Br_{y}\hat{\mathbf{x}}$.
\begin{align}
\mathbf{K}_{e^{+}}&=\int\textrm{d}^{3}r\:[\Psi^{\dagger}_{e}(\mathbf{r})(-i\mathbf{\nabla})\Psi_{e}(\mathbf{r})+e\tilde{\mathbf{A}}]|_{\textrm{positrons}}=\hat{\mathbf{z}}\int\frac{\textrm{d}k_{x}\textrm{d}k_{z}}{(2\pi)^{2}}k_{z}\sum_{\mathcal{S}=\pm1}\sum_{\mathcal{N}=0}b^{e\dagger}_{\mathcal{N},\mathcal{S}}b^{e}_{\mathcal{N},\mathcal{S}}\nonumber\\
&+\int\frac{\textrm{d}k_{x}\textrm{d}k_{z}}{(2\pi)^{2}}k_{z}\sum_{\mathcal{N}=1}\Bigl[2\sqrt{E_{\mathcal{N}}^{e}E_{\mathcal{N}-1}^{e}(E_{\mathcal{N}}^{e}+m_{e})(E_{\mathcal{N}-1}^{e}+m_{e})}\Bigr]^{-1}\Bigl\{\frac{\sqrt{eB(\mathcal{N}-1)}}{2\sqrt{2}}\mathcal{F}_{\mathcal{N}}(eB)\nonumber\\
&\times\Bigl[(\hat{\mathbf{x}}+i\hat{\mathbf{y}})b^{e\dagger}_{(\mathcal{N},+)}b^{e}_{(\mathcal{N}-1,+)}+(\hat{\mathbf{x}}-i\hat{\mathbf{y}})b^{e\dagger}_{(\mathcal{N}-1,+)}b^{e}_{(\mathcal{N},+)}\Bigr]+\frac{\sqrt{eB\mathcal{N}}}{2\sqrt{2}}[\mathcal{F}_{\mathcal{N}}(eB)-4eB]\nonumber\\
&\times\Bigl[(\hat{\mathbf{x}}+i\hat{\mathbf{y}})b^{e\dagger}_{(\mathcal{N},-)}b^{e}_{(\mathcal{N}-1,-)}+(\hat{\mathbf{x}}-i\hat{\mathbf{y}})b^{e\dagger}_{(\mathcal{N}-1,-)}b^{e}_{(\mathcal{N},-)}\Bigr]-k_{z}eB\Bigl[(\hat{\mathbf{x}}+i\hat{\mathbf{y}})b^{e\dagger}_{(\mathcal{N},+)}b^{e}_{(\mathcal{N}-1,-)}\nonumber\\
&+(\hat{\mathbf{x}}-i\hat{\mathbf{y}})b^{e\dagger}_{(\mathcal{N}-1,-)}b^{e}_{(\mathcal{N},+)}\Bigr]\Bigr\},\label{Ke}
\end{align}
where $\mathcal{F}_{\mathcal{N}}(eB)\equiv(E_{\mathcal{N}}^{e}+m_{e}+k_{z})(E_{\mathcal{N}-1}^{e}+m_{e}+k_{z})+(E_{\mathcal{N}}^{e}+m_{e}-k_{z})(E_{\mathcal{N}-1}^{e}+m_{e}-k_{z})+4\mathcal{N}eB$, and the dependence of the creation and annihilation positron operators on $k_{x}$ and $k_{z}$ has been omitted for brevity. 

\section{Dynamical observables in the adiabatic limit}\label{Dyna}
The fact that the external magnetic field $\mathbf{B}$ which enters eq.(\ref{Hmag}) increases adiabatically in time, makes it possible to apply the adiabatic theorem to the computation of the expectation value of any operator which evolves adiabatically with $\mathbf{B}$. Let us consider that $\mathbf{B}(\tau)$ increases slowly in time from $\mathbf{0}$ at $\tau=0$ to $\mathbf{B}_{0}$ at some later time $t_{0}>0$. The expression for the expectation value of any given Schr\"odinger operator $ O $ of the system at time $t>0$ reads,
\begin{equation}
\langle O (t)\rangle=\langle\Phi(0)|\mathbb{U}^{\dagger}(t)\: O \:\mathbb{U}(t)|\Phi(0)\rangle,\label{equO}
\end{equation}
where $|\Phi(0)\rangle$ is the state of the system at $\tau=0$  and $\mathbb{U}(t)=T\:\exp{[-i\int^{t}_{0}H}\:$d$\tau]$ is the time-propagator in Schr\"odinger's representation. In our perturbative approach, we will expand  $\mathbb{U}(t)$ in powers of $W$. In order to set this explicitly, we rather write 
\begin{equation}
\mathbb{U}(t)=\mathbb{U}_{0}(t)\:T\exp{\left[-i\int_{0}^{t}\textrm{d}\tau\:\mathbb{U}_{0}^{\dagger}(\tau)W\mathbb{U}_{0}(\tau)\right]},\label{U1tot}
\end{equation}
where $\mathbb{U}_{0}(t)=\exp{[-i\int^{t}_{0}H_{0}}\:$d$\tau]$. On the other hand, the adiabatic theorem implies that, if the state of the system starts in an eigenstate of $H$ at $\tau=0$, $\mathbf{B}(0)=\mathbf{0}$, it will evolve towards the corresponding eigenstate of the Hamiltonian $H$ at any latter time $t$ at which $\mathbf{B}(t)\neq\mathbf{0}$ can be considered quasi-stationary. Given the one-to-one correspondence between the value of the field $\mathbf{B}$ and the time $t$, hereafter we will replace  the time-dependence of states and operators on $t$ with their dependence on $\mathbf{B}$. In particular, let us define $|\Phi(0)\rangle=|\Phi\rangle_{\mathbf{0}}\equiv|\Omega_{l}\rangle_{\mathbf{0}}\otimes|\phi_{int}^{N}\rangle\otimes[|0,1;\mathbf{0}\rangle+|0,-1;\mathbf{0}\rangle]^{p}_{\mathbf{0}}/2Lm_{p}^{1/2}$ 
as the initial state of the system, where we use the nomenclature of section \ref{hamil} to denote the state of  an unpolarized proton at rest and the leptonic vacuum, at zero magnetic field, with $L\rightarrow\infty$ being a normalization length. The adiabatic theorem implies that, at a later time $t$, this state will evolve towards its corresponding eigenstates of $H(\mathbf{B})$ with a quasi-stationary magnetic field $\mathbf{B}$. Using the nomenclature in section \ref{hamil},
\begin{align}
|\Phi\rangle_{\mathbf{B}}=\mathbb{U}(\mathbf{B})|\Phi\rangle_{\mathbf{0}}\simeq |\Omega_{l}\rangle_{\mathbf{B}}\otimes|\phi_{int}^{0}\rangle\otimes\bigl[\bigl|0,1;\langle\mathbf{P}_{N}\rangle\bigl\rangle+|0,-1;\langle\mathbf{P}_{N}\rangle\bigl\rangle\bigr]^{p}_{\mathbf{B}}/2L\sqrt{E_{0}^{p}}.\label{PB}
\end{align}
 Note that, in order to keep the computation of $\langle O (t)\rangle$ at lowest order in $W$, $|\Phi\rangle_{\mathbf{B}}$ has been approximated to $\mathbb{U}_{0}(\mathbf{B})|\Phi\rangle_{\mathbf{0}}$ but for  the shift $\langle\mathbf{P}_{N}\rangle_{\mathbf{B}}$ in the linear momentum that the nucleon ($N$) acquires as a result of the action of $W$ in eq.(\ref{U1tot}). This way, in our approximation,  $|\Phi\rangle_{\mathbf{B}}$ remains being an eigenstate of $H_{0}(\mathbf{B})$. 

Next, the expectation value $\langle O (t_{0})\rangle=\langle O \rangle_{\mathbf{B}_{0}}$ can be computed out of the integration of the variations $\delta\langle O \rangle_{\mathbf{B}}$ with respect to the adiabatic variations of the magnetic field, $\delta\mathbf{B}$, from $\mathbf{B}=\mathbf{0}$ to $\mathbf{B}_{0}$. In order to compute those variations it is convenient to define a new time-evolution operator which takes account of the adiabatic evolution of the system upon which the operator $ O $ applies. At leading order in $W$ we define
\begin{equation}
\tilde{\mathbb{U}}_{\mathbf{B}}(t)\equiv\tilde{\mathbb{U}}_{\mathbf{B}}^{0}(t)-i\int_{-\infty}^{t}\textrm{d}\tau\:\tilde{\mathbb{U}}_{\mathbf{B}}^{0}(t-\tau)W\tilde{\mathbb{U}}_{\mathbf{B}}^{0}(\tau)e^{\eta\tau},\:\eta\rightarrow0^{+},\label{eta}
\end{equation}
where $\tilde{\mathbb{U}}_{\mathbf{B}}^{0}(\tau)=e^{-iH_{0}(\mathbf{B})\tau}$, for $\mathbf{B}$ stationary. Using this definition we can write the adiabatic variation $\delta\langle O \rangle$ with respect to an adiabatic variation $\delta\mathbf{B}$, at certain value of the magnetic field $\mathbf{B}$, as 
\begin{align}
\delta\langle O \rangle_{\mathbf{B}}=&\Bigl[\:_{\mathbf{0}}\langle\Phi|\frac{\delta}{\delta\mathbf{B}}\Bigl(\mathbb{U}^{\dagger}(\mathbf{B})\tilde{\mathbb{U}}^{\dagger}_{\mathbf{B}}(t)\Bigr)\: O \:\tilde{\mathbb{U}}_{\mathbf{B}}(t)\mathbb{U}(\mathbf{B})|\Phi\rangle_{\mathbf{0}}\nonumber\\
&+\:_{\mathbf{0}}\langle\Phi|\mathbb{U}^{\dagger}(\mathbf{B})\tilde{\mathbb{U}}^{\dagger}_{\mathbf{B}}(t)\: O \:\frac{\delta}{\delta\mathbf{B}}\Bigl(\tilde{\mathbb{U}}_{\mathbf{B}}(t)\mathbb{U}(\mathbf{B})\Bigr)|\Phi\rangle_{\mathbf{0}}\Bigr]\delta\mathbf{B}.\label{dO}
\end{align}
Therefore, in the adiabatic limit, at time $t_{0}$, eq.(\ref{equO}) reads 
\begin{equation}
\langle O (t_{0})\rangle=\int_{\mathbf{0}}^{\mathbf{B}_{0}}\delta\langle O \rangle_{\mathbf{B}}.\label{Otot}
\end{equation}
It is worth noting that the formalism presented here is a generalisation of the so-called Hellmann-Feynman-Pauli theorem which applies to the case that $ O =H_{0}(\mathbf{B})$. That is, to the computation of the energy of a system whose Hamiltonian depends on an adiabatic parameter \cite{Pines}. 

We finalize this section with some comments which are particularly relevant to our calculations. In the first place, we note that since $|\Phi\rangle_{\mathbf{B}}$ is a stationary state throughout the adiabatic variation of $\mathbf{B}$,  eqs.(\ref{eta}) and (\ref{dO}) imply that $\delta\langle O \rangle_{\mathbf{B}}$ can be computed using stationary perturbation theory upon the time-independent Hamiltonian $H_{0}(\mathbf{B})+W$. Further, provided that the operator $ O $ is independent of the adiabatic parameter $\mathbf{B}$, $\langle O (t_{0})\rangle$ can be computed using stationary perturbation theory upon the time-independent Hamiltonian $H_{0}(\mathbf{B}_{0})+W$. That is, if $\delta O /\delta\mathbf{B}=\mathbf{0}$, it holds that, for $t\gtrsim t_{0}$, 
\begin{equation}
\langle O (t)\rangle=\:_{\mathbf{B}_{0}}\langle\Phi|\:\tilde{\mathbb{U}}^{\dagger}_{\mathbf{B_{0}}}(t)\: O \:\tilde{\mathbb{U}}_{\mathbf{B}_{0}}(t)\:|\Phi\rangle_{\mathbf{B}_{0}},\quad t\gtrsim t_{0}.\label{Ototstat}
\end{equation}
In fact, from eq.(\ref{eta}) it is plain that $\langle O (t)\rangle$ in eq.(\ref{Ototstat}) does not depend explicitly on $t$. 

\section{Quadratic fluctuations of leptonic and hadronic currents}\label{AppC}

The hadronic and leptonic currents flanking the energy denominator of the integrands in eqs.(\ref{equationmother}) give rise to quadratic vacuum fluctuations for the case of leptons, and quadratic hadron fluctuations upon the single proton state $|\phi^{p}_{ext}\rangle_{\mathbf{B}_{0}}\equiv\bigl[\bigl|0,1;\mathbf{0}\bigl\rangle+|0,-1;\mathbf{0}\bigl\rangle\bigr]^{p}_{\mathbf{B}_{0}}/2L\sqrt{E_{0}^{p}}$ in  eq.(\ref{equationmother}). We write the equations of these fluctuations as integrals over the momenta of the virtual particles.

\subsection{Leptonic vacuum fluctuations}

The equation for the leptonic current fluctuations in the leptonic vacuum state, $|\Omega_{l}\rangle_{\mathbf{B}_{0}}$, is  
\begin{align}
&_{\mathbf{B}_{0}}\langle\Omega_{l}|J^{l}_{\mu}(\mathbf{r})J^{l\dagger}_{\rho}(\mathbf{r}')|\Omega_{l}\rangle_{\mathbf{B}_{0}}=\int\frac{\textrm{d}^{3}q}{(2\pi)^{3}}\frac{\textrm{d}k_{1}\textrm{d}k_{3}}{(2\pi)^{2}}\frac{e^{i(\mathbf{r}-\mathbf{r}')\cdot(\mathbf{q}+\mathbf{k}')}e^{-i(r_{2}-r_{2}')k_{2}}}{2E_{q}2E^{e}_{\mathcal{N}}}\sum_{\mathcal{N}=0}^{\infty}\sum_{s,r=\pm1}\textrm{Tr}\Bigr\{\bar{V}_{\mathcal{N},s}^{e^{+}}\nonumber\\
&\cdot[\gamma_{\mu}(\mathbb{I}-\gamma_{5})]U^{\nu}_{r}\bar{U}^{\nu}_{r}[\gamma_{\rho}(\mathbb{I}-\gamma_{5})]V_{\mathcal{N},s}^{e^{+}}\Bigl\}
=\int\frac{\textrm{d}^{3}q}{(2\pi)^{3}q}\frac{\textrm{d}k_{x}\textrm{d}k_{z}}{(2\pi)^{2}}e^{i(\mathbf{r}-\mathbf{r}')\cdot(\mathbf{q}+\mathbf{k}')}e^{-i(r_{2}-r_{2}')k_{2}}\Big\{I_{0}(\xi_{e^{+}})\nonumber\\
&\times I_{0}(\xi'_{e^{+}})(1+k_{3}/E_{0}^{e})\Big[q(2\delta_{\mu0}\delta_{\rho0}-g_{\mu\rho}-\delta_{\mu0}\delta_{\rho3}-\delta_{\mu3}\delta_{\rho0}+i\epsilon_{0\mu\rho3})
+ q^{i}(\delta_{\mu0}\delta_{\rho i}+\delta_{\mu i}\delta_{\rho0}\nonumber\\
&+i\epsilon_{0\mu\rho i}-\delta_{\mu i}\delta_{\rho3}-\delta_{\mu3}\delta_{\rho i}-g_{\mu\rho}\delta_{i3}-i\epsilon_{\mu\rho i3})\Big]+I_{0}(\xi_{e^{+}})I_{0}(\xi'_{e^{+}})(1-k_{3}/E_{1}^{e})\Big[q(2\delta_{\mu0}\delta_{\rho0}-g_{\mu\rho}\nonumber\\
&+\delta_{\mu0}\delta_{\rho3}+\delta_{\mu3}\delta_{\rho0}-i\epsilon_{0\mu\rho3})+ q^{i}(\delta_{\mu0}\delta_{\rho i}+\delta_{\mu i}\delta_{\rho0}+i\epsilon_{0\mu\rho i}+\delta_{\mu i}\delta_{\rho3}+\delta_{\mu3}\delta_{\rho i}+g_{\mu\rho}\delta_{i3}+i\epsilon_{\mu\rho i3})\Big]\nonumber\\
&+\sum_{\mathcal{N}=1}^{\infty}I_{\mathcal{N}}(\xi_{e^{+}})I_{\mathcal{N}}(\xi'_{e^{+}})[k_{3}eB_{0} /(E_{\mathcal{N}}^{e})^{3}]\Big[q(2\delta_{\mu0}\delta_{\rho0}-g_{\mu\rho}+\delta_{\mu0}\delta_{\rho3}+\delta_{\mu3}\delta_{\rho0}-i\epsilon_{0\mu\rho3})+ q^{i}(\delta_{\mu0}\delta_{\rho i}\nonumber\\
&+\delta_{\mu i}\delta_{\rho0}+i\epsilon_{0\mu\rho i}+\delta_{\mu i}\delta_{\rho3}+\delta_{\mu3}\delta_{\rho i}+g_{\mu\rho}\delta_{i3}+i\epsilon_{\mu\rho i3})\Big]+2I_{\mathcal{N}}(\xi_{e^{+}})I_{\mathcal{N}}(\xi'_{e^{+}})\Big[q(2\delta_{\mu0}\delta_{\rho0}-g_{\mu\rho})\nonumber\\
&-(k_{3}q/E_{\mathcal{N}}^{e})(\delta_{\mu0}\delta_{\rho3}+\delta_{\mu3}\delta_{\rho0}-i\epsilon_{0\mu\rho3})+ q^{i}(\delta_{\mu0}\delta_{\rho i}+\delta_{\mu i}\delta_{\rho0}+i\epsilon_{0\mu\rho i})-(k_{3}q^{i}/E_{\mathcal{N}}^{e})(\delta_{\mu i}\delta_{\rho3}\nonumber\\
&+\delta_{\mu3}\delta_{\rho i}+g_{\mu\rho}\delta_{i3}+i\epsilon_{\mu\rho i3})\Big] - \sum_{\mathcal{N}=1}^{\infty}\frac{\sqrt{2\mathcal{N}eB_{0}}}{E_{\mathcal{N}}^{e}}[I_{\mathcal{N}}(\xi_{e^{+}})I_{\mathcal{N}-1}(\xi'_{e^{+}})+I_{\mathcal{N}-1}(\xi_{e^{+}})I_{\mathcal{N}}(\xi'_{e^{+}}]\nonumber\\
&\times\Big[q(\delta_{\mu1}\delta_{\rho0}+\delta_{\mu0}\delta_{\rho1}+i\epsilon_{1\mu\rho0})+q^{i}(\delta_{\mu1}\delta_{\rho i}+\delta_{\mu i}\delta_{\rho1}+g_{\mu\rho}\delta_{i1}+i\epsilon_{1\mu\rho0})\Big]
+i \sum_{\mathcal{N}=1}^{\infty}\frac{\sqrt{2\mathcal{N}eB_{0}}}{E_{\mathcal{N}}^{e}}\nonumber\\
&\times[I_{\mathcal{N}}(\xi_{e^{+}})I_{\mathcal{N}-1}(\xi'_{e^{+}})-I_{\mathcal{N}-1}(\xi_{e^{+}})I_{\mathcal{N}}(\xi'_{e^{+}}]\Big[q(\delta_{\mu2}\delta_{\rho0}+\delta_{\mu0}\delta_{\rho2}+i\epsilon_{2\mu\rho0})+q^{i}(\delta_{\mu2}\delta_{\rho i}+\delta_{\mu i}\delta_{\rho2}\nonumber\\
&+g_{\mu\rho}\delta_{i2}+i\epsilon_{2\mu\rho0})\Big]\Big\},\label{SM3}
\end{align}
where greek indices take the values $0,1,2,3$, whereas latin indices take $1,2,3$. The index $0$ corresponds to the time component, while $1,2,3$ refer to the spatial cartesian components $x,y,z$, respectively. In eq.(\ref{SM3}) the factors proportional to $eB_{0}/(E_{\mathcal{N}}^{e})^{2}$ result from the approximation $E_{\mathcal{N}}^{e}/E_{\mathcal{N}+1}^{e}\simeq1-eB_{0}/(E_{\mathcal{N}}^{e})^{2}$. It is only the terms proportional to $(1+k_{3}/E_{0}^{e})$ in the second and third rows that correspond to fluctuations of positrons in the lowest Landau level.

\subsection{Hadronic current fluctuations}

As for the hadronic current fluctuations which enter eq.(\ref{equationmother}), the external proton state is the stationary state $|\phi^{p}_{ext}\rangle_{\mathbf{B}_{0}}$, and the current fluctuations read, up to terms of order $eB_{0}/m_{p}^{2}$,

\begin{align}
&_{\mathbf{B}_{0}}\langle\phi^{p}_{ext}|J^{h}_{\mu}(\mathbf{r})J^{h\dagger}_{\rho}(\mathbf{r}')|\phi^{p}_{ext}\rangle_{\mathbf{B}_{0}}=\int\frac{\textrm{d}^{3}p}{(2\pi)^{3}}\sum_{s,r=\pm1}\frac{e^{i(\mathbf{r}-\mathbf{r}')\cdot\mathbf{p}}}{2L^{2}2E_{r}^{n}2E^{p}_{s}}\textrm{Tr}\{\bar{U}_{0,s}^{p}[\gamma_{\mu}(\mathbb{I}-g_{A}\gamma_{5})]U^{n}_{r}\bar{U}^{n}_{r}\nonumber\\
&\times[\gamma_{\rho}(\mathbb{I}-g_{A}\gamma_{5})]U_{0,s}^{p}\}=\int\frac{\textrm{d}^{3}p}{(2\pi)^{3}}\frac{e^{i(\mathbf{r}-\mathbf{r}')\cdot\mathbf{p}}}{4L{2}}\Big\{4I_{0}(r_{2})I_{0}(r'_{2})\Big[\delta_{\mu0}\delta_{\rho0}+g_{A}^{2}(\delta_{\mu0}\delta_{\rho0}-g_{\mu\rho})\nonumber\\
&+\frac{{p}^{i}}{2m_{n}}[(1+g_{A}^{2})(\delta_{\mu0}\delta_{\rho i}+\delta_{\mu i}\delta_{\rho0})-2g_{A}i\epsilon_{0\mu i\rho}]\Big]-\frac{eB_{0}}{2m_{p}^{2}}[I_{0}(r_{2})I_{0}(r'_{2})+I_{1}(r_{2})I_{1}(r'_{2})]\nonumber\\
&\times\Big[(1-g_{A}^{2})g_{\mu\rho}-2g_{A}(\delta_{\mu3}\delta_{\rho0}+\delta_{\mu0}\delta_{\rho3})+(1+g_{A}^{2})i\epsilon_{0\mu\rho3}+\frac{p^{i}}{m_{n}}[ (1+g_{A}^{2})i\epsilon_{\mu\rho i3} - 2g_{A}(\delta_{\mu3}\delta_{\rho i}\nonumber\\
&+\delta_{\mu i}\delta_{\rho3}+ \delta_{i3}g_{\mu\rho})]\Big]+ \frac{eB_{0}}{2m_{p}^{2}}[I_{1}(r_{2})I_{1}(r'_{2})-I_{0}(r_{2})I_{0}(r'_{2})]\Big[(1+g_{A}^{2})[2\delta_{\mu0}\delta_{\rho0}-g_{\mu\rho}+\frac{p^{i}}{m_{n}}(\delta_{\mu0}\delta_{\rho i}\nonumber\\
&+\delta_{\mu i}\delta_{\rho0})]-(1+g_{A}^{2})i\epsilon_{03\mu\rho}\Big]- \frac{\sqrt{eB_{0}}}{\sqrt{2}m_{p}}[I_{0}(r_{2})I_{1}(r'_{2})+I_{1}(r_{2})I_{0}(r'_{2})]\Big[(1+g_{A}^{2})(\delta_{\mu2}\delta_{\rho0}+\delta_{\mu0}\delta_{\rho2})\nonumber\\
&-2g_{A}i\epsilon_{02\mu\rho}+\frac{p^{i}}{m_{n}}[(1+g_{A}^{2})(\delta_{\mu2}\delta_{\rho i}+\delta_{\mu i}\delta_{\rho2}+ \delta_{i2}g_{\mu\rho}) + 2g_{A}i\epsilon_{i2\mu\rho}]\Big]+4i\frac{\sqrt{eB_{0}}}{\sqrt{2}m_{p}}[I_{0}(r_{2})I_{1}(r'_{2})\nonumber\\
&+I_{1}(r_{2})I_{0}(r'_{2})](1-g_{A}^{2})(\delta_{\mu0}\delta_{\rho1}-\delta_{\mu1}\delta_{\rho0})+\frac{|g_{n}|\mu_{N}B_{0}}{2m_{p}^{2}}I_{0}(r_{2})I_{0}(r'_{2})\Big[(1+g_{A})^{2}(2\delta_{\mu0}\delta_{\rho0}-\delta_{\mu0}\delta_{\rho3}\nonumber\\
&-\delta_{\mu3}\delta_{\rho0})-2g_{A}(1+g_{A})(g_{\mu\rho}+i\epsilon_{03\mu\rho}) + \frac{p_{3}}{2m_{n}}[(1-g_{A}^{2})g_{\mu\rho} + (1+3g_{A}^{2})i \epsilon_{03\mu\rho}+4g_{A}(2\delta_{\mu0}\delta_{\rho0}\nonumber\\
&+\delta_{\mu0}\delta_{\rho3}+\delta_{\mu3}\delta_{\rho0}-g_{\mu\rho})]+ \frac{p_{1}}{2m_{n}}[(g_{A}^{2}-1)(\delta_{\mu2}\delta_{\rho0}-\delta_{\mu0}\delta_{\rho2} - i\epsilon_{01\mu\rho})+(1+g_{A})^{2}(\delta_{\mu0}\delta_{\rho1}\nonumber\\
&+\delta_{\mu1}\delta_{\rho0}+i\epsilon_{01\mu\rho})]+\frac{p_{2}}{2m_{n}}[(g_{A}^{2}-1)(\delta_{\mu1}\delta_{\rho0}-\delta_{\mu0}\delta_{\rho1}-i\epsilon_{02\mu\rho})+(1+g_{A})^{2}(\delta_{\mu0}\delta_{\rho2}+\delta_{\mu2}\delta_{\rho0}\nonumber\\
&+i\epsilon_{02\mu\rho})]\Big]\Big\}.\label{SM5}
\end{align}

\section{Kinetic energy terms in $\mathcal{E}_{\mathcal{N},m}$ and form factors}\label{AppD}

In this appendix we derive the kinetic terms which enter the factor $\mathcal{E}_{\mathcal{N},m}$ in the denominator of the integrand of eq.(\ref{eq:fa:1}), as well as the form factors $\langle\tilde{\phi}^{0}_{int}|\tilde{\phi}_{int}^{m}\rangle$ which appear in that equation. 

\subsection{Kinetic energy terms in $\mathcal{E}_{\mathcal{N},m}$}\label{AppD1}
We follow an approach analogous to that in chapter 3 of ref.~\cite{Milonnibook}, P.W. Milonni, and references therein, which is applied there to the computation of the Lamb-shift. Let us consider the integrand of eq.(\ref{equationmother}), and let us evaluate the energy denominator there for certain intermediate state of positrons, neutrinos and neutrons $|I_{e,\nu,n}\rangle_{\mathbf{B}_{0}}= |\mathcal{N},\mathcal{S}_{e^{+}};k_{x},k_{z}\rangle_{\mathbf{B}_{0}}^{e}\otimes|\mathbf{q},\mathcal{S}_{\nu}\rangle^{\nu}\otimes|\phi_{int}^{m}\rangle\otimes|\mathbf{p},\mathcal{S}\rangle^{n}_{\mathbf{B}_{0}}$,
\begin{equation}
_{\mathbf{B}_{0}}\langle I_{e,\nu,n}|[H_{int}^{N}+H_{D}^{n}+H_{D}^{e}+H_{D}^{\nu}-E_{0}^{p}]|I_{e,\nu,n}\rangle_{\mathbf{B}_{0}}=\langle\phi_{int}^{m}|H^{N}_{int}|\phi_{int}^{m}\rangle+E_{\mathcal{S}}^{n}+E_{\mathcal{N}}^{e}+E_{q}-E_{0}^{p},\label{SM7}
\end{equation}
where the expressions for $\langle\phi_{int}^{m}|H^{N}_{int}|\phi_{int}^{m}\rangle$, $E_{\mathcal{S}}^{n}$, $E_{\mathcal{N}}^{e}$, $E_{q}$ and $E_{0}^{p}$ are given in section \ref{Approach}. Let us substitute the expression in eq.(\ref{SM7}) into the integrand of eq.(\ref{eq:fa:1}), including  the spatial wavefunction of the  $\mathcal{N}^{th}$ positron Landau level, $e^{-\tilde{\xi}^{2}_{e^{+}}/2}H_{\mathcal{N}}(\tilde{\xi}_{e^{+}})$,  and the integrals over the coordinates $x$, $z$, $p_{x}$ and $p_{z}$ of the nucleon center of mass,
\begin{align}
&\int\frac{\textrm{d}x\textrm{d}z\textrm{d}p_{x}\textrm{d}p_{z}}{(2\pi)^{2}}\langle\phi^{0}_{int}|e^{i\tilde{\mathbf{r}}\cdot(\mathbf{p}+\mathbf{k}+\mathbf{q})-ik_{y}\tilde{r}_{y}}e^{-\tilde{\xi}^{2}_{e^{+}}/2}H_{\mathcal{N}}(\tilde{\xi}_{e^{+}})\frac{1}{E^{e}_{\mathcal{N}}+E_{q}+E^{n}_{\mathcal{S}}(\mathbf{p})-E_{0}^{p}+H_{int}^{N}(\mathbf{p}_{\rho};\mathbf{p}_{\lambda})}\nonumber\\
&\times|\phi_{int}^{m}\rangle\hat{\mathbf{z}}\cdot(\mathbf{k}+\mathbf{q})\langle\phi_{int}^{m}|\frac{1}{E^{e}_{\mathcal{N}}+E_{q}+E^{n}_{\mathcal{S}}(\mathbf{p})-E_{0}^{p}+H_{int}^{N}(\mathbf{p}_{\rho};\mathbf{p}_{\lambda})} e^{i\tilde{\mathbf{r}}'\cdot(\mathbf{p}+\mathbf{k}+\mathbf{q})-ik_{y}\tilde{r}_{y}'}e^{-\tilde{\xi}^{'2}_{e^{+}}/2}\nonumber\\
&\times H_{\mathcal{N}}(\tilde{\xi}'_{e^{+}})|\phi_{int}^{0}\rangle,\label{SM8}
\end{align}
where $\tilde{\xi}_{e^{+}}=\sqrt{eB}\tilde{r}_{y}+k_{x}/\sqrt{eB}$, and we have rewritten the coordinate of the active quark in terms of the nucleon center of mass $\mathbf{R}$ and the Jacobi coordinate $\mathbf{r}_{\lambda}$, $\tilde{\mathbf{r}}=\mathbf{R}+(2/3)\mathbf{r}_{\lambda}$, and the internal nucleon Hamiltonian of eq.(\ref{Hint}) as an explicit function of $\mathbf{p}_{\rho}$ and $\mathbf{p}_{\lambda}$. The exponential factors flanking the denominator come from the hadronic and leptonic currents on either side of the integrand in eq.(\ref{equationmother}). Bearing in mind that the relevant contribution of positrons to eq.(\ref{eq:fa:1}) comes from relativistic positrons, the following approximation can be used for Landau levels with $\mathcal{N}\gg1$,
\begin{equation}
e^{-\tilde{\xi}^{2}_{e^{+}}/2}H_{\mathcal{N}}(\tilde{\xi}_{e^{+}})\simeq e^{-\xi^{2}_{e^{+}}/2}H_{\mathcal{N}}(\xi_{e^{+}})\cos{(2r^{y}_{\lambda}\sqrt{2\mathcal{N}eB}/3)}.\nonumber
\end{equation}
Substituting the above expression into eq.(\ref{SM8}), we arrive at
\begin{align}
&\int\frac{\textrm{d}x\textrm{d}z\textrm{d}p_{x}\textrm{d}p_{z}}{(2\pi)^{2}}\langle\phi^{0}_{int}|e^{i\mathbf{R}\cdot(\mathbf{p}+\mathbf{k}+\mathbf{q})-ik_{y}R_{y}}e^{i(2/3)[\mathbf{r}_{\lambda}\cdot(\mathbf{k}+\mathbf{q})-k_{y}r_{\lambda}^{y}]}e^{-\xi^{2}_{e^{+}}/2}H_{\mathcal{N}}(\xi_{e^{+}})\nonumber\\
&\times\cos{(2r^{y}_{\lambda}\sqrt{2\mathcal{N}eB}/3)}\frac{1}{E^{e}_{\mathcal{N}}+E_{q}+E^{n}_{\mathcal{S}}(\mathbf{p})-E_{0}^{p}+H_{int}^{N}(\mathbf{p}_{\rho};\mathbf{p}_{\lambda})}|\phi_{int}^{m}\rangle\hat{\mathbf{z}}\cdot(\mathbf{k}+\mathbf{q})\langle\phi_{int}^{m}|\nonumber\\
&\times \frac{1}{E^{e}_{\mathcal{N}}+E_{q}+E^{n}_{\mathcal{S}}(\mathbf{p})-E_{0}^{p}+H_{int}^{N}(\mathbf{p}_{\rho};\mathbf{p}_{\lambda})}e^{-\xi^{'2}_{e^{+}}/2}H_{\mathcal{N}}(\xi'_{e^{+}})\cos{(2r^{y}_{\lambda}\sqrt{2\mathcal{N}eB}/3)}\nonumber\\
&\times e^{-i\mathbf{R}'\cdot(\mathbf{p}+\mathbf{k}+\mathbf{q})+ik_{y}R'_{y}}e^{-i(2/3)[\mathbf{r}_{\lambda}\cdot(\mathbf{k}+\mathbf{q})-k_{y}r_{\lambda}^{y}]}|\phi_{int}^{0}\rangle.\label{SM9}
\end{align}
The integration in $x$ and $z$ yields the deltas of momentum conservation $(2\pi)^{2}\delta(p_{x}+k_{x}+q_{x})\delta(p_{z}+k_{z}+q_{z})$.  Further, integration in momentum coordinates leads to the following effective replacement in the denominator of eq.(\ref{SM9}),
 \begin{equation}
 E^{n}_{\mathcal{S}}(p_{x},p_{y},p_{z})\rightarrow E^{n}_{\mathcal{S}}(-k_{x}-q_{x},p_{y},-k_{z}-q_{z}).\label{SM10}
 \end{equation} 
 Next, we apply the canonical commutation relations, $[r_{\lambda}^{i},p_{\lambda}^{j}]=i\delta_{ij}$, to move the $r_{\lambda}^{y}$-dependent exponentials to the middle of eq.(\ref{SM9}). Making use of the relationship
 \begin{align}
&\langle\phi^{0}_{int}|e^{i(2/3)[\mathbf{r}_{\lambda}\cdot(\mathbf{k}+\mathbf{q})+r_{\lambda}^{y}(\pm\sqrt{2\mathcal{N}eB}r^{y}_{\lambda}-k_{y})]}\frac{1}{E^{e}_{\mathcal{N}}+E_{q}+E^{n}_{\mathcal{S}}-E_{0}^{p}+H_{int}^{N}(\mathbf{p}_{\rho};\mathbf{p}_{\lambda})}|\phi_{int}^{m}\rangle\nonumber\\
&=\frac{1}{E^{e}_{\mathcal{N}}+E_{q}+E^{n}_{\mathcal{S}}-E_{0}^{p}+\langle\phi^{m}_{int}|H_{int}^{N}[\mathbf{p}_{\rho};\mathbf{p}_{\lambda}-2[\mathbf{k}+\mathbf{q}+\hat{\mathbf{y}}(\pm\sqrt{2\mathcal{N}eB}r^{y}_{\lambda}-k_{y})]/3]|\phi_{int}^{m}\rangle}\nonumber\\&\times\langle\phi^{m}_{int}|e^{i(2/3)[\mathbf{r}_{\lambda}\cdot(\mathbf{k}+\mathbf{q})+r_{\lambda}^{y}(\pm\sqrt{2\mathcal{N}eB}r^{y}_{\lambda}-k_{y})]}|\phi_{int}^{0}\rangle,\label{SM11}
\end{align}
and the replacement in eq.(\ref{SM10}), the energy denominators of eq.(\ref{SM9}) read, for a nonrelativistic neutron,
\begin{align}
 &E_{n}^{\mathcal{S}}(-x_{x}-q_{x},p_{y},-k_{z}-q_{z})+\langle\phi^{m}_{int}|H_{int}^{N}[\mathbf{p}_{\rho};\mathbf{p}_{\lambda}-2[\mathbf{k}+\mathbf{q}+\hat{\mathbf{y}}(\pm\sqrt{2\mathcal{N}eB}r^{y}_{\lambda}-k_{y})]/3]|\phi_{int}^{m}\rangle\nonumber\\
 &+E^{e}_{\mathcal{N}}+E_{q}-E_{0}^{p}=\langle\phi^{m}_{int}|\Bigl[\frac{|\mathbf{k}+\mathbf{q}+(\pm\sqrt{2\mathcal{N}eB}-k_{y})\hat{\mathbf{y}}|^{2}}{2m_{d}}-\frac{\mathbf{p}_{\lambda}}{m_{d}}\cdot[(\pm\sqrt{2\mathcal{N}eB}-k_{y})\hat{\mathbf{y}}\nonumber\\
 &+\mathbf{k}+\mathbf{q}]+V_{\textrm{conf}}^{N}\Bigl]|\phi_{int}^{m}\rangle+E^{e}_{\mathcal{N}}+E_{q}-(\mathcal{S}g_{n}+1)\mu_{N}B+(m_{n}-m_{p})+\frac{p^{2}_{y}}{2m_{n}}.\label{SM12}
 \end{align}
where we have replaced $H_{int}^{N}$ with its expression in eq.(\ref{Hint}). Further, we observe in eq.(\ref{eq:fa:1}) that all the nonvanishing integrals in $y$ and $y'$  yield factors of the form 
\begin{equation}
\int_{-\infty}^{\infty}I_{0,1}(y)I_{\mathcal{N}}(\xi_{e^{+}})e^{iy(p_{y}+q_{y})}\textrm{d}y\propto e^{-[k_{x}^{2}+(p_{y}+q_{y})^{2}]/2eB},\nonumber
\end{equation}
which suppress exponentially the remaining integrands on $k_{x}$, $p_{y}$ and $q_{y}$, for $|k_{x}|>\sqrt{eB}$ and $|p_{y}+q_{y}|>\sqrt{eB}$. In particular, for $eB\ll m^{2}_{d},m^{2}_{n}$, this implies that the terms proportional to $k_{x}/m_{d}$ are negligible and that we can replace $p_{y}^{2}/2m_{n}\rightarrow q_{y}^{2}/2m_{n}$ at the same order of approximation in eq.(\ref{SM12}) . In addition, the terms $\mathbf{p}_{\lambda}\cdot[\mathbf{k}+\mathbf{q}+(\pm\sqrt{2\mathcal{N}eB}-k_{y})\hat{\mathbf{y}}]/m_{d}$ are Doppler shift terms which are negligible in the nonrelativistic limit, and so are the spin-dependent magnetic terms. As a result, the energy denominators of eq.(\ref{SM11}) can be approximated by
 \begin{equation}
 \mathcal{E}_{\mathcal{N},m}^{\pm}\simeq E^{e}_{\mathcal{N}}+E_{q}+m_{n}-m_{p}+(k_{z}^{2}+q^{2}+2\mathcal{N}eB)/2m_{d}+q_{z}k_{z}/m_{d}+m\omega\pm q_{y}\sqrt{2\mathcal{N}eB}/m_{d}.\label{SM13}
 \end{equation} 
As a function of $\mathcal{E}_{\mathcal{N},m}^{\pm}$, eq.(\ref{SM11}) can be written as
\begin{align}
&\left[\frac{\langle\phi^{0}_{int}|e^{i(2/3)[\mathbf{r}_{\lambda}\cdot(\mathbf{k}+\mathbf{q})+r_{\lambda}^{y}(\sqrt{2\mathcal{N}eB}-k_{y})]}|\phi_{int}^{m}\rangle}{\mathcal{E}^{+}_{\mathcal{N},m}}+\frac{\langle\phi^{0}_{int}|e^{i(2/3)[\mathbf{r}_{\lambda}\cdot(\mathbf{k}+\mathbf{q})+r_{\lambda}^{y}(-\sqrt{2\mathcal{N}eB}-k_{y})]}|\phi_{int}^{m}\rangle}{\mathcal{E}^{-}_{\mathcal{N},m}}\right]\nonumber\\
&\times e^{-\xi^{2}_{e^{+}}/2}H_{\mathcal{N}}(\xi_{e^{+}})\frac{e^{i(p_{y}+q_{y})(y-y')}(\mathbf{k}+\mathbf{q})\cdot\hat{\mathbf{z}}}{4}e^{-\xi^{'2}_{e^{+}}/2}H_{\mathcal{N}}(\xi'_{e^{+}})\label{SM14}\\
&\times\left[\frac{\langle\phi^{m}_{int}|e^{-i(2/3)[\mathbf{r}_{\lambda}\cdot(\mathbf{k}+\mathbf{q})+r_{\lambda}^{y}(\sqrt{2\mathcal{N}eB}-k_{y})]}|\phi_{int}^{0}\rangle}{\mathcal{E}^{+}_{\mathcal{N},m}}+\frac{\langle\phi^{m}_{int}|e^{-i(2/3)[\mathbf{r}_{\lambda}\cdot(\mathbf{k}+\mathbf{q})+r_{\lambda}^{y}(-\sqrt{2\mathcal{N}eB}-k_{y})]}|\phi_{int}^{0}\rangle}{\mathcal{E}^{-}_{\mathcal{N},m}}\right].\nonumber
\end{align}
In this equation we identify the form factors between the wavefunctions of the internal nucleon states $0$ and $m$,
\begin{equation}
\langle\phi^{0}_{int}|e^{\pm i(2/3)[\mathbf{r}_{\lambda}\cdot(\mathbf{k}+\mathbf{q})+r_{\lambda}^{y}(\pm\sqrt{2\mathcal{N}eB}-k_{y})]}|\phi_{int}^{m}\rangle,\label{formfactors}\\
\end{equation}
which have been denoted by $\langle\tilde{\phi}^{0}_{int}|\tilde{\phi}_{int}^{m}\rangle$ in eq.(\ref{eq:fa:1}) of Sec.\ref{Adiabatic}.   Finally, approximating $\mathcal{E}_{\mathcal{N},m}^{+}\simeq\mathcal{E}_{\mathcal{N},m}^{-}\simeq\mathcal{E}_{\mathcal{N},m}=E^{e}_{\mathcal{N}}+E_{q}+m_{n}-m_{p}+(k_{z}^{2}+q^{2}+2\mathcal{N}eB)/2m_{d}+q_{z}k_{z}/m_{d}+m\omega$,  eq.(\ref{SM11}) reads
\begin{align}
e^{i(p_{y}+q_{y})(y-y')}e^{-(\xi^{2}_{e^{+}}+\xi^{'2}_{e^{+}})/2}H_{\mathcal{N}}(\xi_{e^{+}})H_{\mathcal{N}}(\xi'_{e^{+}})\Big|\langle\tilde{\phi}^{0}_{int}|\tilde{\phi}_{int}^{m}\rangle\Big|^{2}\frac{k_{z}+q_{z}}{\mathcal{E}^{2}_{\mathcal{N},m}},\label{SM15}
\end{align} 
as it appears in eq.(\ref{eq:fa:1}).

\subsection{Form factors $\langle\tilde{\phi}^{0}_{int}|\tilde{\phi}_{int}^{m}\rangle$}\label{AppD2}

Regarding the form factors of the internal states and the contributions of these states to the energy denominator $\mathcal{E}_{\mathcal{N},m}$,  we first note that the $\mathbf{r}_{\lambda}$-dependent and $\mathbf{r}_{\rho}$-dependent parts of the Hamiltonian $H_{int}^{N}$ in eq.(\ref{Hint}) are isotropic harmonic oscillators of frequency $\omega$ and effective masses $2m_{d}/3$ and $m_{d}/2$, respectively. As for the sum over internal nucleon states in eq.(\ref{eq:fa:1}), the states $|\phi^{m}_{int}\rangle$ are composed of the tensor product of the ground state of the oscillator in $\mathbf{r}_{\rho}$, and all the states of the oscillator in $\mathbf{r}_{\lambda}$. As for the wavefunction of the non-degenerate ground state, it reads
\begin{equation}
\varphi_{int}^{0}(\mathbf{r}_{\lambda},\mathbf{r}_{\rho})=\langle \mathbf{r}_{\lambda},\mathbf{r}_{\rho}|\phi_{int}^{0}\rangle=\frac{\beta^{3}}{3^{3/4}\pi^{3/2}}e^{-\beta^{2}(r_{\lambda}^{2}/3+r_{\rho}^{2}/4)},\quad\beta=\sqrt{\omega m_{d}},\label{phinot}
\end{equation}
while the excited eigenstates with energy $m\omega$, $m\in\mathbb{N}$, are degenerate with degree of degeneracy $g_{m}=(m+1)(m+2)/2$ and wavefunctions
\begin{align}
\varphi_{int}^{m_{x},m_{y},m_{z}}(\mathbf{r}_{\lambda},\mathbf{r}_{\rho})&=\frac{\varphi_{int}^{0}(\mathbf{r}_{\lambda},\mathbf{r}_{\rho})}{\sqrt{2^{m}m_{x}!m_{y}!m_{z}!}}H_{m_{x}}(\sqrt{2/3}\beta r_{\lambda}^{x})H_{m_{y}}(\sqrt{2/3}\beta r_{\lambda}^{y})H_{m_{z}}(\sqrt{2/3}\beta r_{\lambda}^{z}),\nonumber\\m&=m_{x}+m_{y}+m_{z}.
\end{align}
Inserting the above expression in eq.(\ref{formfactors}) for the form factors, and squaring, we end up with
\begin{equation}
|\langle\tilde{\phi}^{0}_{int}|\tilde{\phi}_{int}^{m}\rangle|^{2}=\frac{(3/4)^{m}}{m_{x}!m_{y}!m_{z}!}(\kappa_{x}/\beta)^{2m_{x}}(\kappa_{y}/\beta)^{2m_{y}}(\kappa_{z}/\beta)^{2m_{z}}e^{-3(\kappa_{x}^{2}+\kappa_{y\pm}^{2}+\kappa_{z}^{2})+/4\beta^{2}},\label{formy}%\quad m= m_{x}+m_{y}+m_{z},
\end{equation}
where $\kappa_{x}=2(k_{x}+q_{x})/3$, $\kappa_{z}=2(k_{z}+q_{z})/3$, $\kappa_{y\pm}=2(q_{y}\pm\sqrt{2\mathcal{N}eB})/3$. On the other hand, the energy of the excited sates $m\omega$ becomes rapidly relativistic, since $\omega\sim m_{d}$. Therefore, in an effective manner, the combination $|\langle\tilde{\phi}^{0}_{int}|\tilde{\phi}_{int}^{m}\rangle|^{2}/\mathcal{E}_{\mathcal{N},m}^{2}$ in the integrand of eq.(\ref{eq:fa:1}) scales rapidly as 
\begin{equation}
|\langle\tilde{\phi}^{0}_{int}|\tilde{\phi}_{int}^{m}\rangle|^{2}/\mathcal{E}_{\mathcal{N},m}^{2}\approx\frac{(3/4)^{m}}{m_{x}!m_{y}!m_{z}!m^{2}\omega^{2}}(\kappa_{x}/\beta)^{2m_{x}}(\kappa_{y}/\beta)^{2m_{y}}(\kappa_{z}/\beta)^{2m_{z}}e^{-3(\kappa_{x}^{2}+\kappa_{y\pm}^{2}+\kappa_{z}^{2})+/4\beta^{2}}.\label{formy1}%,\quad m= m_{x}+m_{y}+m_{z}
\end{equation}
In the first place, we note from eqs.(\ref{formy}) and (\ref{formy1}) that the combination $|\langle\tilde{\phi}^{0}_{int}|\tilde{\phi}_{int}^{m}\rangle|^{2}/\mathcal{E}_{\mathcal{N},m}^{2}$ suppresses exponentially the integrand of the momentum integrals in eq.(\ref{eq:fa:1}) for all $m$. It provides this way an effective cutoff at $q,k,\sqrt{2\mathcal{N}eB}\approx\beta\sim m_{d}$, which makes the integrals convergent. Second, the form factor of the excited state $m$ provides a factor which scales as $\sim(k^{2}+q^{2}+2\mathcal{N}eB)^{m}/m_{d}^{2m}$, which is a relativistic correction $\forall$ $m\geq1$ that lies beyond the non-relativistic approximation assumed for the internal hadron dynamics. Therefore, the contribution of all the excited states must be discarded by consistency. Nonetheless, one must bear in mind that their contribution would be just one order higher than the leading order nonrelativistic terms included in eq.(\ref{f1b}). With this proviso, we are just left with the contribution of the internal nucleon ground state whose form factor reads
\begin{align}
\langle\phi^{0}_{int}|&e^{\pm i(2/3)[\mathbf{r}_{\lambda}\cdot(\mathbf{k}+\mathbf{q})+r_{\lambda}^{y}(\pm\sqrt{2\mathcal{N}eB}-k_{y})]}|\phi_{int}^{0}\rangle=\int\textrm{d}^{3}r_{\rho}\int\textrm{d}^{3}r_{\lambda}e^{\pm i(2/3)[\mathbf{r}_{\lambda}\cdot(\mathbf{k}+\mathbf{q})+r_{\lambda}^{y}(\pm\sqrt{2\mathcal{N}eB}-k_{y})]}\nonumber\\
&\times\frac{\beta^{6}}{(\sqrt{3}\pi)^{3}}e^{-\beta^{2}(2r_{\lambda}^{2}/3+r_{\rho}^{2}/2)}=e^{-[(k_{x}+q_{x})^{2}+(k_{z}+q_{z})^{2}+q_{y}^{2}+2\mathcal{N}eB]/6\beta^{2}}e^{\mp q_{y}\sqrt{2\mathcal{N}eB}/3\beta^{2}}.\label{SM16}
\end{align}
Substituting this formula into eq.(\ref{SM14}) we end up with
\begin{align}
&(1/2)e^{i(p_{y}+q_{y})(y-y')}e^{-(\xi^{2}_{e^{+}}+(\xi')^{2}_{e^{+}})/2}H_{\mathcal{N}}(\xi_{e^{+}})H_{\mathcal{N}}(\xi'_{e^{+}})e^{-[(k_{x}+q_{x})^{2}+(k_{z}+q_{z})^{2}+q_{y}^{2}+2\mathcal{N}eB]/3\beta^{2}}\nonumber\\
&\times(\mathbf{k}+\mathbf{q})\left[\frac{e^{-2q_{y}\sqrt{2\mathcal{N}eB}/3\beta^{2}}}{(\mathcal{E}^{+}_{\mathcal{N}})^{2}}+\frac{1}{\mathcal{E}^{+}_{\mathcal{N}}\mathcal{E}^{-}_{\mathcal{N}}}\right],\nonumber
\end{align}
where it is assumed that $\sqrt{2\mathcal{N}eB}$ takes both positive and negative values. 

\section{Magnetic nature of the proton kinetic energy}\label{AppE}

In this appendix we demonstrate in detail the assertions made in section \ref{AdiaEnergy} regarding the magnetic nature of the proton kinetic energy as well as its origin as a Doppler-shift correction to the EW self-interaction.

\begin{figure}[h!]
\centering
\includegraphics[height=6.3cm,width=6.cm,clip]{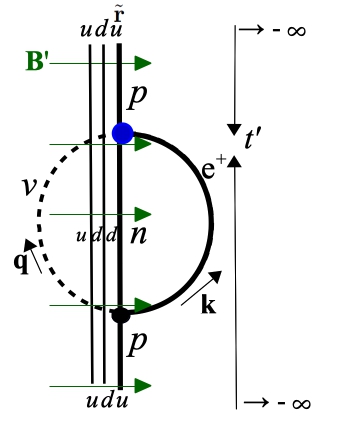}
\caption{Diagrammatic representation of $\langle W\rangle_{\mathbf{B}'}$ for a quasi-stationary magnetic field $\mathbf{B}'$, according to eq.(\ref{SM19}). The calculation is performed in the adiabatic limit, where the initial time is taken to $-\infty$ according to eq.(\ref{eta}), and the operator $W$, represented with a blue circle, applies at the observation time $t'$.}\label{SMfig}
\end{figure}

Let us consider first the variation of the total energy of the proton with respect to variations of the magnetic field strength, $\delta B$, at certain value of the field $\mathbf{B}$,
\begin{equation}
\langle\delta H/\delta B\rangle_{\mathbf{B}}=\langle\delta[H_{0}+W]/\delta B\rangle_{\mathbf{B}}=\langle\delta[H_{mag}^{p}+H_{mag}^{n}+H_{mag}^{e}]/\delta B\rangle_{\mathbf{B}},\nonumber\qquad
\end{equation}
where the total Hamiltonian is evaluated in the state of the system at a certain value of the field $\mathbf{B}$, which is denoted by $\rangle_{\mathbf{B}}$. Taking into account that $B$ is an adiabatic parameter, $\delta B$ can be factored out of the expectation values of the above equation,
\begin{equation}
\langle\delta H/\delta B\rangle_{\mathbf{B}}=\langle\delta[H_{mag}^{p}+H_{mag}^{n}+H_{mag}^{e}]/\delta B\rangle_{\mathbf{B}}=\delta\left\langle\sum_{\alpha}H_{D}^{\alpha}+W+\delta T\right\rangle_{\mathbf{B}}/\delta B.\label{SM17}
\end{equation}
Next, we will show that the variation of the kinetic energy of the proton is indeed part of the variation of the magnetic energy, 
 since it is one of the terms within $\delta\langle W\rangle_{\mathbf{B}}/\delta B$. That is, we will prove that 
\begin{align}
&\frac{\delta\langle\mathbf{P}_{N}\rangle_{\mathbf{B}}^{2}}{2m_{p}\delta B}\textrm{ is part of }\delta\langle W\rangle_{\mathbf{B}}/\delta B,\textrm{ which is part of }\langle\delta H/\delta B\rangle_{\mathbf{B}}=\left\langle\frac{\delta\left(H_{mag}^{p}+H_{mag}^{n}+H_{mag}^{e}\right)}{\delta B}\right\rangle_{\mathbf{B}},\nonumber\\
&\textrm{from which it holds that }\frac{\delta\langle\mathbf{P}_{N}\rangle_{\mathbf{B}}^{2}}{2m_{p}\delta B}\textrm{ is part of }\left\langle \delta\left(H_{mag}^{p}+H_{mag}^{n}+H_{mag}^{e}\right)/\delta B\right\rangle_{\mathbf{B}}.\label{SM18}
\end{align}
Here $\langle\mathbf{P}_{N}\rangle_{\mathbf{B}}$ is the momentum of the proton at magnetic field $\mathbf{B}$, computed according to eq.(\ref{equationmother}), which is linear in $\mathbf{B}$. 
Let us start with the expression for $\langle W\rangle_{\mathbf{B}'}$ at a quasi-stationary field $\mathbf{B}(t')\equiv\mathbf{B}'=\mathbf{B}+\delta\mathbf{B}$ whose diagrammatic representation is that of figure~\ref{SMfig}, at $\mathcal{O}(W^{2})$,
\begin{align}
\langle W\rangle_{\mathbf{B}'}&=_{\mathbf{B}'}\langle\tilde{\Phi}|\tilde{\mathbb{U}}^{\dagger}_{\mathbf{B}'}(t')\:W\:\tilde{\mathbb{U}}_{\mathbf{B}'}(t')|\tilde{\Phi}\rangle_{\mathbf{B}'}\nonumber\\
&=-\frac{G_{F}^{2}}{2}\textrm{Re}\int\textrm{d}^{3}R\textrm{d}^{3}R'\:_{\mathbf{B}'}\langle\tilde{\Phi}|J_{h}^{\mu}(\mathbf{R},\tilde{\mathbf{r}})J^{l}_{\mu}(\tilde{\mathbf{r}})\frac{1}{H_{0}-E_{0}^{p}}
J^{l\dagger}_{\rho}(\tilde{\mathbf{r}}')|J_{h}^{\rho\dagger}(\mathbf{R}',\tilde{\mathbf{r}}')|\tilde{\Phi}\rangle_{\mathbf{B}'},\label{SM19}
\end{align}
where $|\tilde{\Phi}\rangle_{\mathbf{B}'}=|\Omega_{l}\rangle_{\mathbf{B}'}\otimes|\phi_{int}^{0}\rangle\otimes\big[\big|0,1;\langle\mathbf{P}_{N}\rangle_{\mathbf{B}}\big\rangle+\big|0,-1;\langle\mathbf{P}_{N}\rangle_{\mathbf{B}}\big\rangle\big]^{p}_{\mathbf{B}'}/2L\sqrt{E^{p}_{0}}$ is the normalized state of the system in a magnetic field $\mathbf{B}'$, with momentum $\langle\mathbf{P}_{N}\rangle_{\mathbf{B}}$ along $\mathbf{B}$, and $E_{0}^{p}$ incorporates its kinetic energy $|\langle\mathbf{P}_{N}\rangle_{\mathbf{B}}|^{2}/2m_{p}$.  Denoting by $\mathbf{R}$ the center of mass vector of the proton, we can write instead, $|\tilde{\Phi}\rangle_{\mathbf{B}'}=|\Omega_{l}\rangle_{\mathbf{B}'}\otimes|\phi_{int}^{0}\rangle\otimes e^{i\langle\mathbf{P}_{N}\rangle_{\mathbf{B}}\cdot\mathbf{R}}\big[|0,1;\mathbf{0}\rangle+|0,-1;\mathbf{0}\rangle\big]^{p}_{\mathbf{B}'}/2L\sqrt{E^{p}_{0}}$.  Note also that $\mathbf{B}'$ and $\mathbf{B}$ differ by $\delta\mathbf{B}$. This will allow us to identify $\delta\langle W\rangle_{\mathbf{B}}$ from the relationship $\langle W\rangle_{\mathbf{B}'}=\langle W\rangle_{\mathbf{B}}+\delta\langle W\rangle_{\mathbf{B}}$.
In appendix \ref{AppD} we computed the energy factor $\mathcal{E}_{\mathcal{N},m}$ starting with the expression in eq.(\ref{SM9}).  In the present case, that equation must incorporate the factor $e^{-i\langle\mathbf{P}_{N}\rangle_{\mathbf{B}}\cdot\mathbf{R}}$ of $\:_{\mathbf{B}'}\langle\tilde{\Phi}|$ on the left, and $e^{i\langle\mathbf{P}_{N}\rangle_{\mathbf{B}}\cdot\mathbf{R}'}$ on the right of the integrand,
\begin{align}
&\int\frac{\textrm{d}x\textrm{d}z\textrm{d}p_{x}\textrm{d}p_{z}}{(2\pi)^{2}}\langle\phi^{0}_{int}|e^{i\mathbf{R}\cdot(\mathbf{p}+\mathbf{k}+\mathbf{q}-\langle\mathbf{P}_{N}\rangle_{\mathbf{B}})-ik_{y}R_{y}}e^{i(2/3)[\mathbf{r}_{\lambda}\cdot(\mathbf{k}+\mathbf{q})-k_{y}r_{\lambda}^{y}]}e^{-\xi^{2}_{e^{+}}/2}H_{\mathcal{N}}(\xi_{e^{+}})\nonumber\\
&\times\cos{(2r^{y}_{\lambda}\sqrt{2\mathcal{N}eB}/3)}|\phi_{int}^{m}\rangle\frac{1}{E^{e}_{\mathcal{N}}+E_{q}+E^{n}_{\mathcal{S}}(\mathbf{p})-E_{0}^{p}+H_{int}^{N}(\mathbf{p}_{\rho};\mathbf{p}_{\lambda})}\langle\phi_{int}^{m}| e^{-\xi^{'2}_{e^{+}}/2}H_{\mathcal{N}}(\xi'_{e^{+}})\nonumber\\
&\times\cos{(2r^{y}_{\lambda}\sqrt{2\mathcal{N}eB}/3)} e^{-i\mathbf{R}'\cdot(\mathbf{p}+\mathbf{k}+\mathbf{q}+\langle\mathbf{P}_{N}\rangle_{\mathbf{B}})+ik_{y}R'_{y}}e^{-i(2/3)[\mathbf{r}_{\lambda}\cdot(\mathbf{k}+\mathbf{q})-k_{y}r_{\lambda}^{y}]}|\phi_{int}^{0}\rangle.\label{SM20}
\end{align}
In contrast to eq.(\ref{SM9}), the integration in $z$ of eq.(\ref{SM20}) yields now $2\pi\delta(p_{z}+q_{z}+k_{z}-\langle P_{N}\rangle_{\mathbf{B}})$, from which the energy factor of eq.(\ref{SM13}) picks up an additional term, 
\begin{equation}
\mathcal{E}^{\pm}_{\mathcal{N},m}\rightarrow \mathcal{E}^{\pm}_{\mathcal{N},m}-(\mathbf{k}+\mathbf{q})\cdot\langle\mathbf{P}_{N}\rangle_{\mathbf{B}}/m_{p}.\quad\nonumber
\end{equation}
The additional term is the energy associated to the Doppler shift of the virtual positrons and neutrinos which interact with the proton in motion. Following steps analogous to those in appendix \ref{AppD}, eq.(\ref{SM20}) can be written as
\begin{align}
\frac{e^{i(p_{y}+q_{y})(y-y')}e^{-(\xi^{2}_{e^{+}}+(\xi')^{2}_{e^{+}})/2}H_{\mathcal{N}}(\xi_{e^{+}})H_{\mathcal{N}}(\xi'_{e^{+}})\Big|\langle\tilde{\phi}^{0}_{int}|\tilde{\phi}_{int}^{m}\rangle\Big|^{2}}{\mathcal{E}_{\mathcal{N},m}-(\mathbf{k}+\mathbf{q})\cdot\langle\mathbf{P}_{N}\rangle_{\mathbf{B}}/m_{p}}.\label{SM21}
\end{align} 
Finally, expanding the above equation up to leading order in $(\mathbf{k}+\mathbf{q})\cdot\langle\mathbf{P}_{N}\rangle_{\mathbf{B}}/m_{p}\mathcal{E}_{\mathcal{N},m}$, eq.(\ref{SM21}) is approximately equal to
\begin{align}
e^{i(p_{y}+q_{y})(y-y')}e^{-(\xi^{2}_{e^{+}}+(\xi')^{2}_{e^{+}})/2}H_{\mathcal{N}}(\xi_{e^{+}})H_{\mathcal{N}}(\xi'_{e^{+}})\Big|\langle\tilde{\phi}^{0}_{int}|\tilde{\phi}_{int}^{m}\rangle\Big|^{2}\left[\frac{1}{\mathcal{E}_{\mathcal{N},m}}+\frac{(\mathbf{k}+\mathbf{q})\cdot\langle\mathbf{P}_{N}\rangle_{\mathbf{B}}/m_{p}}{\mathcal{E}^{2}_{\mathcal{N},m}}\right].\label{SM22}
\end{align}
The first term does not depend on the proton momentum, whereas the second term is exactly eq.(\ref{SM15}) found in appendix \ref{AppD} for the Casimir momentum times $\langle\mathbf{P}_{N}\rangle_{\mathbf{B}}/m_{p}$. Putting all these results together, eq.(\ref{SM19}) yields, up to terms of order $(\mathbf{k}+\mathbf{q})\cdot\langle\mathbf{P}_{N}\rangle_{\mathbf{B}}/m_{p}\mathcal{E}_{\mathcal{N},m}$ in the integrand,
\begin{align}
\langle&W\rangle_{\mathbf{B}'}=-\frac{G_{F}^{2}}{2}\textrm{Re}\int\textrm{d}^{3}R\textrm{d}^{3}R'\:_{\mathbf{B}'}\langle\Phi|J_{h}^{\mu}(\mathbf{R},\tilde{\mathbf{r}})J^{l}_{\mu}(\tilde{\mathbf{r}})\frac{1}{H_{0}-E_{0}^{p}}
J^{\rho\dagger}_{\mu}(\tilde{\mathbf{r}}')J_{h}^{\rho\dagger}(\mathbf{R}',\tilde{\mathbf{r}}')|\Phi\rangle_{\mathbf{B}'}\nonumber\\
&-\frac{G_{F}^{2}}{2}\textrm{Re}\int\textrm{d}^{3}R\textrm{d}^{3}R'\:_{\mathbf{B}'}\langle\Phi|J_{h}^{\mu}(\mathbf{R},\tilde{\mathbf{r}})J^{l}_{\mu}(\tilde{\mathbf{r}})\frac{(\mathbf{K}_{e}+\mathbf{Q}_{\nu})\cdot\langle\mathbf{P}_{N}\rangle_{B}/m_{p}}{[H_{0}-E_{0}^{p}]^{2}}
J^{\rho\dagger}_{\mu}(\tilde{\mathbf{r}}')J_{h}^{\rho\dagger}(\mathbf{R}',\tilde{\mathbf{r}}')|\Phi\rangle_{\mathbf{B}'},\label{SM23}
\end{align}
where $|\Phi\rangle_{\mathbf{B}'}=|\Omega_{l}\rangle_{\mathbf{B}'}\otimes|\phi_{int}^{0}\rangle\otimes\big[\big|0,1;\mathbf{0}\big\rangle+\big|0,-1;\mathbf{0}\big\rangle\big]^{p}_{\mathbf{B}'}/2L\sqrt{E^{p}_{0}}$ is here the state of the system in a magnetic field $\mathbf{B}'$, where the proton presents zero momentum along $\mathbf{B}$. The first term of eq.(\ref{SM23}) is the leading order EW self-energy of a proton in the presence of a constant and uniform magnetic field $\mathbf{B}'$, while the second term is the leading order Doppler-shift correction. By direct comparison with eq.(\ref{equationmother}), the second term is readily identifiable with $-\langle\mathbf{K}_{e^{+}}+\mathbf{Q}_{\nu}\rangle_{\mathbf{B}'}\cdot\langle\mathbf{P}_{N}\rangle_{\mathbf{B}}/m_{p}$. As argued in section \ref{Adiabatic}, conservation of total momentum along $\mathbf{B}'$ implies $-\langle\mathbf{K}_{e^{+}}+\mathbf{Q}_{\nu}\rangle_{\mathbf{B}'}=\langle\mathbf{P}_{N}\rangle_{\mathbf{B}'}$. Lastly, since $\mathbf{B}'=\mathbf{B}+\delta\mathbf{B}$, we can write the nucleon momentum at $\mathbf{B}'$ as $\langle\mathbf{P}_{N}\rangle_{\mathbf{B}'}=\langle\mathbf{P}_{N}\rangle_{\mathbf{B}}+\delta\langle\mathbf{P}_{N}\rangle_{\mathbf{B}}$, from which we obtain that 
\begin{equation}
\frac{\langle\mathbf{P}_{N}\rangle_{\mathbf{B}'}\cdot\langle\mathbf{P}_{N}\rangle_{\mathbf{B}}}{m_{p}}=\frac{\langle\mathbf{P}_{N}\rangle_{\mathbf{B}}^{2}}{m_{p}}+\frac{\delta\langle\mathbf{P}_{N}\rangle_{\mathbf{B}}\cdot\langle\mathbf{P}_{N}\rangle_{\mathbf{B}}}{m_{p}}\textrm{ is part of }\langle W\rangle_{\mathbf{B}'}=\langle W\rangle_{\mathbf{B}}+\delta\langle W\rangle_{\mathbf{B}}.\label{SM24}
\end{equation}
Identifying $\delta\langle\mathbf{P}_{N}\rangle_{\mathbf{B}}\cdot\langle\mathbf{P}_{N}\rangle_{\mathbf{B}}/m_{p}$ as part of $\delta\langle W\rangle_{\mathbf{B}}$ in eq.(\ref{SM24}), 
we conclude the proof of the relationship (\ref{SM18}) and confirm the assertions in section \ref{AdiaEnergy}.

\section{Some integrals and sums of infinite series}\label{AppF}

The integrals in eq.(\ref{eq:fa:1}) involve a number of technical issues. Here we summarize the most relevant ones.

As for the integrals involving $I_{\mathcal{N}}$ functions, the following identities hold.
\begin{align}
&\int_{-\infty}^{\infty}\textrm{d}y\textrm{d}y'\int_{-\infty}^{\infty}\frac{\textrm{d}p_{y}\textrm{d}k_{x}}{(2\pi)^{2}}e^{ip_{y}(y-y')}I_{\mathcal{N}}(y)I_{\mathcal{N}}(y')I_{\mathcal{N}'}(\xi_{e^{+}})I_{\mathcal{N}'}(\xi'_{e^{+}})=eB/2\pi,\nonumber\\
&\int_{-\infty}^{\infty}\textrm{d}y\textrm{d}y'\int_{-\infty}^{\infty}\frac{\textrm{d}p_{y}\textrm{d}k_{x}}{2(2\pi)^{2}}e^{ip_{y}(y-y')}k_{x}\sqrt{2\mathcal{N}eB}I_{0}(y)I_{0}(y')[I_{\mathcal{N}}(\xi_{e^{+}})I_{\mathcal{N}-1}(\xi'_{e^{+}})\nonumber\\
&+I_{\mathcal{N}}(\xi'_{e^{+}})I_{\mathcal{N}-1}(\xi_{e^{+}})]=\mathcal{N}e^{2}B^{2}/\pi.\nonumber
\end{align}

Regarding the sums over infinite series, considering that $eB\ll m_{d}^{2}$ and that the range of $k_{z}$ for which the integrands yield relevant contributions to the momentum integrals is such that $|k_{z}|\lessapprox m_{d}$, we perform the passage to the continuum by means of the following substitutions,
\begin{equation}
E^{2}_{\mathcal{N}B}\equiv2\mathcal{N}eB,\qquad\sum_{\mathcal{N}=0}^{\infty}f(E_{\mathcal{N}B})\rightarrow\int_{0}^{\infty}\frac{E_{\mathcal{N}B}\textrm{d}E_{\mathcal{N}B}}{eB}f(E_{\mathcal{N}B}),\quad\textrm{for any function }f\textrm{ of }E_{\mathcal{N}B}.\nonumber
\end{equation}


\begin{thebibliography}{105}
\bibitem{Miltonbook} K.A. Milton, \emph{The Casimir Effect: Physical Manifestations of Zero-point Energy}, World Sci., Singapore (2001).
\bibitem{Milonnibook}  P.W. Milonni,  \textit{The Quantum Vacuum}, Academic Press, San Diego (1994).
\bibitem{Weinberg} S.  Weinberg,  \textit{The cosmological constant problem}, Rev.  Mod. Phys. \textbf{61} (1989) 1.
\bibitem{Jaffe} R.L. Jaffe, \textit{Casimir effect and the quantum vacuum}, Phys. Rev. D \textbf{72} (2005) 021301(R).
\bibitem{Casimir} H.B.G. Casimir, Proc. K. Ned. Akad. Wet. \textbf{51}, 793 (1948); E. M. Lifschitz, \textit{The theory of molecular attractive forces between solids}, Sov. Phys. \textbf{2} (1956) 73; S.K. Lamoreaux,  \textit{Demonstration of the Casimir Force in the 0.6 to 6$\mu$m Range}, Phys. Rev. Lett. \textbf{78} (1997) 5, Erratum: Phys. Rev. Lett. \textbf{81} (1998) 5475. 
\bibitem{vdW} F. London, \textit{Zur Theorie und Systematik der Molekularkr\"{a}fte}, Z. Phys. \textbf{63} (1930) 245; H.B.G. Casimir  and D. Polder, \textit{The Influence of Retardation on the London-van der Waals Forces}, \emph{Phys. Rev.} \textbf{73} (1948) 360.
 \bibitem{Bethe} W.E. Lamb and R.C. Retherford, \textit{Fine Structure of the Hydrogen Atom by a Microwave Method}, Phys. Rev. \textbf{72} (1947) 241; H. A. Bethe, \textit{The Electromagnetic Shift of Energy Levels}, Phys. Rev. \textbf{72} (1947) 339.
 \bibitem{cavitychromo} A. Chodos, R.L. Jaffe, K. Johnson, C.B Thorn and V.F. Weisskopf, \textit{New Extended Model of Hadrons}, Phys. Rev. D \textbf{9} (1974) 3471;  P. Hasenfratz and J. Kuti, \textit{The quark bag model}, Phys. Rep. \textbf{40} (1978) 75. 
 \bibitem{Cherenkov} M.N. Chernodub, V.A. Goy, A.V. Molochkov, and H.H. Nguyen, \textit{Casimir Effect in Yang-Mills Theory in $D=2+1$}, Phys. Rev. Lett. \textbf{121} (2018) 191601.
 \bibitem{PRL99071601(2007)Durrer} R. Durrer and M. Russer, \textit{Dynamical Casimir Effect in Braneworlds}, Phys. Rev. Lett. \textbf{99} (2007) 071601.
\bibitem{Feigel} A. Feigel, \textit{Quantum Vacuum Contribution to the Momentum of Dielectric Media}, Phys. Rev. Lett. \textbf{92} (2004) 020404.
\bibitem{Croze}  B.A. van Tiggelen, G.L.J.A. Rikken and V. Krsti\'c, \textit{Momentum Transfer from Quantum Vacuum to Magnetoelectric Matter},  Phys. Rev. Lett. \textbf{96} (2006) 130402; B.A. van Tiggelen, \textit{Zero-point momentum in complex media}, Eur. Phys. J. D \textbf{47} (2008) 261;  B.A. van Tiggelen, S. Kawka and G.L.J.A. Rikken, \textit{QED Corrections to the Electromagnetic Abraham Force. Casimir Momentum of the Hydrogen Atom?}, Eur. Phys. J. D \textbf{66}  (2012) 272; O.A. Croze, \textit{Alternative derivation of the Feigel effect and call for its experimental verification}, Proc. R. Soc. A \textbf{468} (2012) 429. 
\bibitem{PRLDonaire} M. Donaire, B.A. van Tiggelen and G.L.J.A. Rikken, \textit{Casimir Momentum of a Chiral Molecule in a Magnetic Field}, Phys. Rev. Lett. {\bf 111} (2013) 143602.
\bibitem{JPCMDonaire} M. Donaire, B.A. van Tiggelen and G.L.J.A. Rikken, \textit{Transfer of linear momentum from the quantum vacuum to a magnetochiral molecule}, \emph{J. Phys.: Condens. Matter}  {\bf 27} (2015) 214002.
\bibitem{Wu} C.S. Wu, E. Ambler, R.W. Hayward, D.D. Hoppes, and R.P. Hudson, \textit{Experimental Test of Parity Conservation in Beta Decay}, Phys. Rev. 
 \textbf{105} (1957) 1413.
\bibitem{Karliner} M. Karliner and H.J. Lipkin, \textit{New quark relations for hadron masses and magnetic moments: A challenge for explanation from QCD}, Phys. Lett. B \textbf{650} (2007) 185.
\bibitem{Scadron} M.D. Scadron, R. Delbourgo, and G. Rupp, \textit{Constituent Quark Masses and the Electroweak Standard Model}, J. Phys. G: Nucl. Part. Phys. \textbf{32} (2006) 736.
 \bibitem{Mao} G.J. Mao, A. Iwamoto and Z.X.Li, \textit{A Study of Neutron Star Structure in Strong Magnetic Fields that includes Anomalous Magnetic Moments}, Chin. J. Astron. Astrophys. \textbf{3} (2003) 359; H. Wen, L.S. Kisslinger, W. Greiner and J.G. Mao, \textit{Neutron spin polarization in strong magnetic fields}, Int. J. Mod. Phys. E \textbf{14} (2005) 1197.
\bibitem{Peskin} M.E. Peskin and D.V. Schroeder, \textit{An Introduction to Quantum Field Theory}, Westview Press, Chicago (1995); C. Itzykson and J.B. Zuber, \textit{Quantum Field Theory}, McGraw-Hill Inc. (1980); J. J. Sakurai, \textit{Advanced Quantum Mechanics}, Addison-Wesley, Boston (1994).
\bibitem{Bhattacharya} K. Bhattacharya and P.B. Pal,  Pramana - J. Phys.  \textbf{62} (2004) 1041; K. Bhattacharya, \textit{Solution of the Dirac equation in presence of an uniform magnetic field}, arXiv:0705.4275 [hep-th] (2007).
\bibitem{Bander} M. Bander and H.R. Rubinstein, \textit{Proton Beta decay in large magnetic fields}, Phys. Lett. B \textbf{311} (1993) 187.
 \bibitem{Broderick} A. Broderick, M. Prakash and J.M. Lattimer, \textit{The Equation of State of Neutron Star Matter in Strong Magnetic Fields}, ApJ \textbf{537} (2000) 351.
\bibitem{Coince}  L. Conci and M. Traini,  \textit{Quark Momentum Distribution in Nucleons}, Few-Body Systems , \textbf{8} (1990)  123.
\bibitem{IsgurKarl1978}  N. Isgur and G. Karl,  \textit{P-wave baryons in the quark model}, Phys. Rev. D \textbf{18} (1978) 4187.
\bibitem{IsgurKarl1979}  N. Isgur and G. Karl,  \textit{Positive-parity excited baryons in a quark model with hyperfine interactions}, Phys. Rev. D \textbf{19} (1979) 2653.
\bibitem{deRujula}  A. de R\'ujula, H. Georgi and S.L. Glashow,  \textit{Hadron masses in a gauge theory}, Phys. Rev. D \textbf{12} (1975)147.
\bibitem{Isgur}  N. Isgur,  \textit{Mesonlike baryons and the spin-orbit puzzle}, Phys. Rev. D\textbf{62} (2000) 014025.
\bibitem{FeynmanGellmann} R.P. Feynman and M. Gell-Mann,  \textit{Theory of the Fermi Interaction},  Phys. Rev. \textbf{109} (1958) 193; 
E.C.G. Sudarshan and R. Marshak,  \textit{Chirality invariance and the Universal Fermi Interaction},  Phys. Rev. \textbf{109} (1958) 1860.
 \bibitem{betadecay} J.D. Jackson, S.B. Treiman and H.W. Wyld. Jr., \textit{Possible Test of Time Reversal Invariance in Beta Decay}, Phys. Rev. \textbf{106}, 517 (1957); J.S. Nico, \textit{Neutron Beta Decay}, J. Phys. G: Nucl. Part. Phys. \textbf{36} (2009) 104001; M.S. Sozzi, \textit{Discrete Symmetries and CP Violation. From Experiment to Theory}, Oxford University Press Inc., New York (2008).
\bibitem{QuarkModels} D. Flamm and F. Schoberl, \textit{Introduction to the Quark Model of Elementary Particles}, Gordon \& Breach Science Pub. (1983); R.F. \'Alvarez-Estrada, F. Fern\'andez, J.L. S\'anchez-G\'omez and V. Vento, \textit{Models of Hadron Structure Based on Quantum Chromodynamics}, Springer-Verlag, Berlin (1986); A.J.G. Hey and R.L. Kelly, \textit{Baryon Spectroscopy}, Phys. Rep. \textbf{96} (1983) 71;   A. Chodos, R.L. Jaffe, K. Johnson and C.B Thorn, \textit{Baryon Structure in the Bag Theory}, Phys. Rev. D \textbf{10} (1974) 2599.
\bibitem{Shulga} S.G. Shul'ga, and T.P. Il'icheva, \textit{Quasi-potential Approach to the Three-quark Nucleon Wave Function}, Russ. Phys. J.  \textbf{47} (2004) 1242.
\bibitem{Khalilov2005}  V.R. Khalilov,  \textit{Electroweak Nucleon Decays in a Superstrong Magnetic Field}, Theor. Math. Phys. \textbf{145} (2005) 1462.
\bibitem{spinrelax}  D.B. Melrose and V.V. Zheleznyakov,  \textit{Quantum Theory of Cyclotron Emission and the X-ray Line in Her X-1}, Astron. Astrophys. \textbf{95} (1981) 86; A.Y. Pothekin and D. Lai,  \textit{Statistical Equilibrium and Ion Cyclotron Absorption/Emission in Strongly Magnetized Plasmas}, Mon. Not. R. Astron. Soc. \textbf{376} (2007) 793.
\bibitem{NMR} C. P. Slichter, \textit{Principles of Magnetic Resonance}, Springer, Berlin (1990).
\bibitem{PRL106_253001(2011)} S. Ulmer, C.C. Rodegheri, K. Blaum, H. Kracke, A. Mooser, W. Quint, and J. Walz, \textit{Observation of Spin Flips with a Single Trapped Proton}, Phys. Rev. Lett. \textbf{106} (2011) 253001.
\bibitem{Pines} D. Pines and P. Nozieres, \emph{The Theory of Quantum Liquids}, Benjamin, New York (1966), pg. 295-7.


\end{thebibliography}
\end{document}